\documentclass{article}
\usepackage{cite}
\usepackage{graphicx}
\usepackage{latexsym}
\usepackage{amsmath}
\usepackage{amssymb}
\usepackage{amsfonts}
\usepackage{amsxtra}

\allowdisplaybreaks
\newcommand{\sss}[1]{{\scriptscriptstyle #1}}

\newcommand{\Ll}{\ensuremath{\mathcal{L}}}

\newsavebox{\piracbox}
\savebox{\piracbox}{$\not\! p$}
\newcommand{\pirac}{\usebox{\piracbox}}
\newsavebox{\ptiracbox}
\savebox{\ptiracbox}{$\not\! p_\sss{T}$}
\newcommand{\ptirac}{\usebox{\ptiracbox}}
\newsavebox{\qiracbox}
\savebox{\qiracbox}{$\not\! q$}

\newcommand{\tr}{\mathrm{Tr}}
\newcommand{\Tr}[1]{\mathrm{Tr}\left( #1 \right)}
\newcommand{\diag}{\mathrm{diag}}

\newcommand{\mm}{\widetilde{m}}
\newcommand{\mb}{m}
\newcommand{\Mb}{M}
\newcommand{\Mbnt}{{M_9}}
\newcommand{\Mv}{{M_V}}

\newcommand{\T}{\mathcal{T}}

\newcommand{\F}{\mathcal{F}}
\renewcommand{\S}{\mathcal{S}}
\newcommand{\A}{\mathcal{A}}
\renewcommand{\P}{\mathcal{P}}

\newcommand{\N}{\mathcal{N}}
\newcommand{\R}{\mathcal{R}}
\newcommand{\U}{\mathcal{U}}
\newcommand{\X}{\mathcal{X}}
\newcommand{\uav}{\langle u\rangle}
\newcommand{\uwd}{\Delta u}
\newcommand{\tav}{\langle t\rangle}
\newcommand{\twd}{\Delta t}

\renewcommand{\Re}{\ensuremath{\mathrm{Re}}}
\renewcommand{\Im}{\ensuremath{\mathrm{Im}}}

\newcommand{\umax}{u_\sss{\mathrm{max}}}
\newcommand{\umin}{u_\sss{\mathrm{min}}}

\newcommand{\tmax}{t_\sss{\mathrm{max}}}
\newcommand{\tmin}{t_\sss{\mathrm{min}}}
\newcommand{\pw}[3]{\ud{f_{#1\!\!}}{#2}{#3}}
\newenvironment{sarray}{\renewcommand{\arraycolsep}{0pt}
  
  \begin{array}}{\end{array}} 
\newcommand{\uud}[4]{#1
  \begin{sarray}{ccc} {\scriptstyle #2} & {\scriptstyle #3} & \\ &  &
    {\scriptstyle #4} \end{sarray}} 
\newcommand{\udu}[4]{#1
  \begin{sarray}{ccc} {\scriptstyle #2} &  & {\scriptstyle #4}\\ &
    {\scriptstyle #3} &  \end{sarray}} 
\newcommand{\dud}[4]{#1
  \begin{sarray}{ccc} & {\scriptstyle #3}  & \\ {\scriptstyle #2} & &
    {\scriptstyle #4}  \end{sarray}} 
\newcommand{\udd}[4]{#1
  \begin{sarray}{ccc} {\scriptstyle #2} & & \\ & {\scriptstyle #3} &
    {\scriptstyle #4}  \end{sarray}} 
\newcommand{\ud}[3]{#1
  \begin{sarray}{cc}{\scriptstyle #2} &  \\ & {\scriptstyle #3}
  \end{sarray}} 
\title{Polarized $pK^-$ scattering in Unitary Baryon Chiral
  Perturbation Theory}
\author{Antonio O.\ Bouzas \thanks{E-mail:
    abouzas@mda.cinvestav.mx}\\\small Departamento de F\'{\i}sica
  Aplicada, CINVESTAV-IPN \\\small Carretera Antigua a Progreso Km.\
  6, Apdo.\ Postal 73 ``Cordemex''\\\small
  M\'erida 97310, Yucat\'an, M\'exico}
\date{}
\begin{document}
\maketitle
\begin{abstract}
We study $pK^-$ scattering in the energy range from threshold
through the $\Lambda(1520)$ peak in UBChPT, taking into account
$\mathcal{O}(q)$ vertices from meson-baryon contact interactions and
$s$- and $u$-channel ground-state baryon exchange, $s$- and
$u$-channel decuplet- and nonet-baryon exchange and $t$-channel
vector-meson exchange, as well as $\mathcal{O}(q^2)$ flavor-breaking
vertices.  Detailed fits to data are presented, including a
substantial body of differential cross-section data with meson
momentum $q_\mathrm{lab} > 300$ MeV not considered in previous
treatments.  
\end{abstract}

\section{Introduction}
\label{sec:intro}

The low energy dynamics of hadrons is succesfully described by Chiral
Perturbation Theory (ChPT), the effective field theory of meson
interactions (see, for example, \cite{ber07,bor07,sch05} for recent
reviews and \cite{wei96,don94} for textbook expositions).  The
effective chiral framework can also be extended to the one-baryon
sector, where a fully relativistic Baryon Chiral Perturbation Theory
(BChPT) has been formulated, describing baryon interactions at low
energies (recent reviews are given in \cite{ber07a,sch05}).  The
domain of applicability of BChPT is limited to energies near the
reaction threshold at which meson momenta are much smaller than the
chiral symmetry breaking scale.  In the case of $N\overline{K}$
scattering, however, the strong coupling among the different open
channels renders the theory inapplicable even at threshold.  

Those limitations of BChPT have motivated the introduction of
unitarization techniques to extend its phenomenological scope to
higher energies in $N\pi$ and $NK^+$ processes, and to the $S=-1$
sector.  Unitary coupled-channels techniques based on
Lippmann--Schwinger or Bethe--Salpeter equations have been succesfully
applied to the study of $N\overline{K}$ and other meson--baryon
processes, even at relatively high energies
\cite{sie88,kai95,kai97,ose98,krp1,krp2,lut00,bor02}.  A unitarization
method dealing directly with the chiral effective theory $T$-matrix
was introduced in \cite{mei00,oll01}, drawing elements from the N/D
method of \cite{che60} and from analogous approaches in the meson
sector \cite{oll99a,oll99}.  This Unitary Baryon Chiral Perturbation
Theory (UBChPT) has been shown to give good descriptions of
cross-section data in $N\overline{K}$ processes
\cite{oll01,jid02,jid03,oll06z}, and in $N\pi$ scattering beyond the
$\Delta$ resonance peak \cite{mei00,bou08}.

In this paper we study $pK^-$ scattering in the energy range from
threshold through the $\Lambda(1520)$ peak, corresponding to
laboratory-frame incident-meson momentum $0\leq q_\mathrm{lab}\lesssim
600$ MeV, in UBChPT.  Specifically, with tree-level BChPT partial
waves as input for the unitarization approach of \cite{mei00, oll01},
we obtain the unitarized partial waves needed to compute physical
observables such as total and differential cross sections and spin
asymmetries, subsequently fit to experimental data.  We take into
account in our tree-level amplitudes $\mathcal{O}(q)$
vertices\footnote{ $\mathcal{O}(q^n)$ denotes a generic quantity of
  chiral order $n$, with $q$ a nominally small quantity such as a
  meson momentum or mass.} from meson-baryon contact interactions and
$s$- and $u$-channel ground-state baryon exchange, $s$- and
$u$-channel decuplet- and nonet-baryon exchange and $t$-channel
vector-meson exchange, as well as $\mathcal{O}(q^2)$ flavor-breaking
vertices.  We include $S$, $P$ and $D$ partial waves in our
computations, the contribution from $F$ and higher waves being
negligibly small at the energies considered here.  A global fit to
$pK^-$ data over a similar energy range, also including $N\pi$ and
$pK^+$ data, was carried out in \cite{lut00} in the unitarization
framework of the Bethe--Salpeter equation.  Unlike \cite{lut00},
however, we explicitly take into account vector-meson exchange, whose
contribution is expected to be significant in this context and, as
discussed below, plays a crucial role in the fits to data presented
here.  Furthermore, we include in our fits a substantial body of
differential cross-section data with meson momentum $q_\mathrm{lab} >
300$ MeV not considered in previous chiral-theoretic treatments.

There is by now overwhelming evidence that the nonets of $J^P=1/2^-$
and $3/2^-$ baryon resonances \cite{PDG} are generated by chiral
coupled-channels dynamics.  The resonance $\Lambda(1405)$ has been
shown to be dynamically generated in \cite{ose98,lut00,oll01,jid03}.
Similarly, the resonances $\Lambda(1670)$ and $\Sigma(1620)$ are
dynamically generated in UBChPT in \cite{oll06z,ose02,sar05}.  Further
examples of the description of nonet baryons as dynamical resonances
are $N(1520)$ \cite{sar05,lut99}, $N(1535)$ \cite{kai95,lut99,ino02},
$\Lambda(1520)$ \cite{sar05}.  There is a profuse literature on
dynamical generation of resonances in UBChPT that we cannot describe
in detail here; recent reviews are given in \cite{lut05,ram07,ose09}.
It should be mentioned that a ``static,'' quark-level component of the
$1/2^-$ and $3/2^-$ baryon nonets cannot be ruled out at present.  In
fact, the existence of such a static component has been advocated,
e.g., in \cite{jid08} in the case of $N(1535)$ and in \cite{jid08a}
for $\Lambda(1520)$.  In this paper we treat the $3/2^-$ baryon nonet
as static, described by an explicit field in the chiral Lagrangian,
whereas the $1/2^-$ nonet is dynamically generated.  We remark that no
global fit to $pK^-$ scattering data including a dynamic $3/2^-$
baryon nonet has been given yet.  We consider the present treatment 
a necessary previous step.

The paper is organized as follows.  In sect.\ \ref{sec:treel} we
discuss the computation of tree-level amplitudes for baryon--meson
scattering in BChPT and their associated partial waves, providing
explicit expressions for those amplitudes and partial waves that have
not been given in the previous literature.  In sect.\ \ref{sec:untr}
we briefly discuss the unitarization procedure \cite{mei00, oll01}
applied in UBChPT.  Our results are presented in sect.\ \ref{sec:res},
where we describe our fitting procedure and best-fit parameters, and
detailedly confront computed observables with experimental data.
Sect.\ \ref{sec:finrem} contains some final remarks.  We gather
technical material, needed in sect.\ \ref{sec:treel}, in the
appendices at the end of the paper.

\section{Tree-level partial waves}
\label{sec:treel}

The ground-state meson and baryon octets are described by standard
\cite{don94} traceless $3\times 3$ complex matrix fields $\phi$ and
$B$, resp., with $\phi$ hermitian.  We use the physical flavor
basis
\begin{equation}
  \label{eq:cartanweyl}
  \begin{gathered}
    \beta^1 = \frac{1}{\sqrt{2}} \left(\lambda^1 + i \lambda^2\right),
    \quad
    \beta^2 = \beta^{1\dagger}~,
    \quad
  \beta^3 = \lambda^3~,\\
  \beta^4 = \frac{1}{\sqrt{2}} \left(\lambda^4 + i \lambda^5\right),
  \quad
  \beta^5 = \beta^{4\dagger}~,\quad  
  \beta^6 = \frac{1}{\sqrt{2}} \left(\lambda^6 + i \lambda^7\right),\quad
  \beta^7 = \beta^{6\dagger}~,\quad  
  \beta^8 = \lambda^8~,
  \end{gathered}
\end{equation}
where $\lambda^a$ are SU(3) Gell-Mann matrices.  The $SU(3)$ algebra
in this basis is described in sect.\ 2 of \cite{bou08}.

Defining the $T$-matrix as $S=I + i (2\pi)^4 \delta(P_f-P_i) T$, the
scattering amplitudes are given by $T$-matrix elements
$\ud{\T}{ab}{a'b'}(s,u;\sigma,\sigma') \equiv \langle
B_{a'}(p',\sigma') M_{b'}(q') | T | B^{a}(p,\sigma) M^b(q) \rangle$ as
functions of the Mandelstam invariants $s=(p+q)^2$, $u = (p-q')^2$ and
the spin variables.  The center-of-mass frame (CMF) partial waves
$\pw{\ell\pm}{ab}{a'b'}$ corresponding to $j=\ell\pm 1/2$ are defined
as,
\begin{equation}
  \label{eq:pw}
  \begin{split}
  \ud{\T}{ab}{a'b'}(s,u;\sigma,\sigma') &= \sum_{\ell = 0}^\infty
  \left\{ 
  \left((\ell+1) \pw{\ell+}{ab}{a'b'} + \ell \pw{\ell-}{ab}{a'b'}
  \right) 
  P_\ell(\widehat{p}\cdot \widehat{p}') \chi'^\dagger_{\sigma'} \cdot
  \chi_{\sigma} \right.\\
  &+ \left. i \left(\pw{\ell+}{ab}{a'b'} - \pw{\ell-}{ab}{a'b'}
  \right) P'_\ell(\widehat{p}\cdot \widehat{p}') \chi'^\dagger_{\sigma'} \cdot
  (\vec{\sigma} \cdot (\widehat{p}\wedge \widehat{p}'))\chi_{\sigma}  \right\}~,
  \end{split}
\end{equation}
with $\widehat{p}\cdot \widehat{p}' = \cos\theta_{CM}$, $P_\ell$ and $P'_\ell$
the Legendre polynomial of order $\ell$ and its derivative, and
$\chi_\sigma$, $\chi'_{\sigma'}$ 2-component spinors for the initial
and final baryon, resp.  

\subsection{Partial waves from octet-baryon exchange}
\label{sec:pwoct}

The Lagrangian of fully relativistic Baryon Chiral Perturbation Theory
(BChPT) is written as a sum $\Ll = \Ll_M + \Ll_{MB}$ of a purely
mesonic Lagrangian $\Ll_M$ and a meson--baryon one $\Ll_{MB}$.  The
mesonic Lagrangian to $\mathcal{O}(q^4)$ was first obtained in
\cite{gas84,gas85a}.  The meson--baryon Lagrangian $\Ll_{MB}$ has been
given to $\mathcal{O}(q^3)$ in the three-flavor case in \cite{oll06a}
(see also \cite{kra90,bor97,fri04}), and in \cite{gas88} for two
flavors.  The tree-level amplitudes from $\Ll$ for meson-baryon
scattering have been given in \cite{oll01} (see also
\cite{jid02,bou08a}).  The associated CM-frame partial waves have been
given in full detail in \cite{bou08}, so there is no need to repeat
them here.  We emphasize that the partial waves of \cite{bou08}, as
well as those given below, are valid only within the physical region
for the process being considered, away from which appropriate
analytic continuation is necessary.

\subsection{Partial waves from decuplet-baryon exchange}
\label{sec:pwdec}

For decuplet-baryon exchange diagrams at tree level, our starting
point is the relativistic Lagrangian for $\Delta$-$N$-$\pi$
interaction from \cite{ber95} (see also, e.g.,
\cite{ben89,ols78,pas07}). At leading chiral order the transition from
two to three flavors amounts to inserting in the amplitudes the flavor
factors for the coupling of two octets and a decuplet.  The tree-level
scattering amplitudes so obtained differ from those for $N\pi$
scattering \cite{mei00} only in their flavor coefficients.  The
CMF partial waves for tree-level $s$-channel decuplet exchange
given in \cite{bou08} are used here without modification.  We discuss
here $u$-channel decuplet-baryon exchange, which was not given
explicitly in \cite{bou08}, and below similar results are presented
for $s$- and $u$-channel nonet-baryon and $t$-channel
vector-meson exchange amplitudes.

The tree-level contribution of $u$-channel decuplet exchange to the
$T$-matrix can be parameterized as,
\begin{equation}
  \label{eq:dec-u-amp}
    \ud{\T_{(u,\mathrm{dec})}}{ab}{a'b'} = \frac{9}{8} g_{10}^2 \frac{(D+F)^2}{f^2}
    \sum_{C=1}^{10}
    \frac{\dud{\S}{a'}{b}{C}\,\udd{\S}{a}{b'}{C}}{u-M_C^2+i0}
    \overline{u}'
    \left(\ud{\widehat{\Gamma}_{(u,\mathrm{dec})0}}{ab}{a'b'C}
    + \ud{\widehat{\Gamma}_{(u,\mathrm{dec})1}}{ab}{a'b'C} \ptirac\right)u~,
\end{equation}
where $g_{10}$ is a coupling constant allowing for departures from the
SU(6) symmetric case $g_{10}=1$, and $u$ and $u'$ are the Dirac spinors
for the initial and final baryon. The flavor coefficients attached
to each vertex in (\ref{eq:dec-u-amp}) are given by,
\begin{equation}
  \label{eq:decflav}
  \dud{\S}{a'}{b}{C} = \frac{1}{2} \varepsilon_{ilm} (\beta^\dagger_{a'})_{lj}
  (\beta^b)_{mk} (T_C)_{ijk}~,
  \quad
  \udd{\S}{a}{b'}{C} = \frac{1}{2} \varepsilon_{ilm} (\beta^a)_{jl}
  (\beta^\dagger_{b'})_{km} (T_C)_{ijk}~,
\end{equation}
with repeated indices $i$, $j$, \ldots, summed from 1 to 3, and the
matrices $\beta$ from (\ref{eq:cartanweyl}).  $T_C$ in
(\ref{eq:decflav}) are a standard basis for the decuplet
representation space (as given, e.g., in eq.\ (9) of \cite{leb94}).
In order to fully specify the reduced amplitudes in
(\ref{eq:dec-u-amp}) we expand them as, 
\begin{equation}
  \label{eq:red-dec-u-amp}
  \ud{\widehat{\Gamma}_{(u,\mathrm{dec})0}}{ab}{a'b'C} = 
\ud{\widehat{\Gamma}_{(u,\mathrm{dec})0.0}}{ab}{a'b'C} +
\ud{\widehat{\Gamma}_{(u,\mathrm{dec})0.1}}{ab}{a'b'C}\, u +
\ud{\widehat{\Gamma}_{(u,\mathrm{dec})0.2}}{ab}{a'b'C}\, u^2 +
\ud{\widehat{\Gamma}_{(u,\mathrm{dec})0.3}}{ab}{a'b'C}\, t~,
\end{equation}
with the coefficients on the r.h.s.\ independent of $u$ and $t$, and
similarly $\ud{\widehat{\Gamma}_{(u,\mathrm{dec})1}}{ab}{a'b'C}$.  This
expansion is also needed to compute the partial waves associated to the
amplitude (\ref{eq:dec-u-amp}).
Explicit calculation yields,
\begin{subequations}
  \label{eq:red-dec-u-amp-explicit}
\begin{equation}
  \label{eq:red-dec-u-amp-explicit-a}
  \begin{aligned}
\ud{\widehat{\Gamma}_{(u,\mathrm{dec})0.0}}{ab}{a'b'C} &=
\frac{1}{3}(\mb_{a'}\mm_b^2+\mb_a\mm_{b'}^2)(1-\kappa) +
\frac{1}{6}(\mb_{a'}^3+\mb_a^3) (1+2\kappa)\\
&+\frac{1}{3}(\mb_{a}\mb_{a'}^2+\mb_a^2\mb_{a'})(1+2\kappa^2)
 + \frac{1}{2}(\mb_{a'}\mm_{b'}^2+\mb_a\mm_{b}^2)
 \\
&
+\frac{\Mb_C}{3}(1-2\kappa+4\kappa^2)(\mb_{a'}^2+\mb_{a}\mb_{a'}+\mb_{a}^2)
+\frac{\Mb_C}{2}(\mm_{b'}^2+\mm_{b}^2)\\ 
&+\frac{1}{6\Mb_C}\left(\rule{0pt}{12pt}\mm_{b'}^2(\mb_{a'}^2-\mm_{b}^2)+\mm_{b}^2(\mb_{a}^2-\mm_{b'}^2)+ 
\mb_{a'}\mb_{a}(\mb_{a'}^2+\mb_{a}^2-\mm_{b'}^2-\mm_{b}^2) \right) \\
&-\frac{1}{6\Mb_C^2}(\mb_{a}+\mb_{a'})(\mb_{a}^2-\mm_{b'}^2)(\mb_{a'}^2-\mm_{b'}^2)~,\\
\ud{\widehat{\Gamma}_{(u,\mathrm{dec})0.1}}{ab}{a'b'C} &=
-\frac{1}{6}(\mb_{a'}+\mb_{a})(1+2\kappa)
-\frac{1}{3\Mb_C} \mb_{a}\mb_{a'} (1-2\kappa+4\kappa^2)
-\frac{1}{6\Mb_C} (\mm_{b'}^2 + \mm_b^2)
\\
&+\frac{2}{3\Mb_C} (\mb_{a'}^2+\mb_{a}^2) \kappa (1-2\kappa)
 - \frac{\Mb_C}{3} (1-2\kappa+4\kappa^2)
+\frac{1}{6\Mb_C^2}\left[\left(\rule{0pt}{12pt}\mb_{a'}(\mb_{a'}^2-\mm_{b}^2)
    \right.\right.\\ 
&
+\left.\left. \mb_{a}(\mb_{a}^2-\mm_{b'}^2)\rule{0pt}{12pt} \right)
(1-2\kappa)+ \mb_{a}\mb_{a'}(\mb_{a}+\mb_{a'})(1-4\kappa^2) 
- (\mb_{a'}\mm_{b'}^2 + \mb_{a}\mm_{b}^2)\right]~,\\
\ud{\widehat{\Gamma}_{(u,\mathrm{dec})0.2}}{ab}{a'b'C} &=
-\frac{2}{3\Mb_C}\kappa(1-2\kappa) - 
\frac{1}{6\Mb_C^2}(1-2\kappa)(\mb_{a}+\mb_{a'})~, \\
\ud{\widehat{\Gamma}_{(u,\mathrm{dec})0.3}}{ab}{a'b'C} &=
-\frac{1}{2}(\mb_{a}+\mb_{a'}+\Mb_C)~, 
  \end{aligned}
\end{equation}
and
\begin{equation}
  \label{eq:red-dec-u-amp-explicit-b}
  \begin{aligned}
\ud{\widehat{\Gamma}_{(u,\mathrm{dec})1.0}}{ab}{a'b'C} &=
-\frac{1}{3}(\mm_b^2+\mm_{b'}^2)(1-\kappa) -
\frac{1}{6}(\mb_{a'}^2+\mb_a^2) (1+2\kappa) -\frac{1}{3}\mb_{a}\mb_{a'}(1+2\kappa^2) \\
&-\frac{\Mb_C}{3} (\mb_{a'}+\mb_{a})(1-2\kappa+4\kappa^2)
+\frac{1}{6\Mb_C}(\mm_{b'}^2\mb_{a'}+\mm_{b}^2\mb_a -\mb_{a}\mb_{a'}^2-\mb_{a}^2\mb_{a'}) \\
&+\frac{1}{6\Mb_C^2}(\mm_{b}^2\mm_{b'}^2-\mm_{b'}^2\mb_{a'}^2-\mm_{b}^2\mb_{a}^2+\mb_{a}^2\mb_{a'}^2)~,\\
\ud{\widehat{\Gamma}_{(u,\mathrm{dec})1.1}}{ab}{a'b'C} &=
\frac{2}{3} \kappa (1-\kappa)
+\frac{1}{6\Mb_C} (\mb_{a} + \mb_{a'})(1-4\kappa+8\kappa^2)\\
&+\frac{1}{6\Mb_C^2}\left[(\mm_{b}^2+\mm_{b'}^2-\mb_{a}^2-\mb_{a'}^2)
(1-2\kappa)+ \mb_{a}\mb_{a'} 4\kappa^2\right]~,\\
\ud{\widehat{\Gamma}_{(u,\mathrm{dec})1.2}}{ab}{a'b'C} &=
\frac{1}{6\Mb_C^2}(1-2\kappa)^2~, \\
\ud{\widehat{\Gamma}_{(u,\mathrm{dec})1.3}}{ab}{a'b'C} &=
\frac{1}{2}~.
  \end{aligned}
\end{equation}
\end{subequations}
In this equation $\kappa=Z+1/2$, with $Z$ the off-shell parameter
entering the Lagrangian for the Rarita--Schwinger decuplet field.%
\footnote{The parameter $Z$ has been shown to be redundant in
  \cite{kre09}.  A formulation without off-shell parameter is given,
  e.g., in \cite{pas07}.  We retain the formulation with an off-shell
  parameter here, to make use of the results of
  \cite{mei00,oll01,bou08}.}  With these coefficients the amplitude
(\ref{eq:dec-u-amp}) is completely specified.  From
(\ref{eq:dec-u-amp}) it is immediate that partial waves for
$u$-channel decuplet exchange take the form
\begin{equation}
  \label{eq:dec-u-pwvs}
    \ud{f_{(u,\mathrm{dec})\ell\pm}}{ab}{a'b'} = \frac{9}{8} g_{10}^2 \frac{(D+F)^2}{f^2}
    \sum_{C=1}^{10}
    \dud{\S}{a'}{b}{C}\,\udd{\S}{a}{b'}{C}
   \ud{\widehat{f}_{(u,\mathrm{dec})\ell\pm}}{ab}{a'b'C}~.
    \end{equation}
In order to express the reduced partial waves
$\ud{\widehat{f}_{(u,\mathrm{dec})\ell\pm}}{ab}{a'b'}$ in terms of the
coefficients (\ref{eq:red-dec-u-amp-explicit-a}),
(\ref{eq:red-dec-u-amp-explicit-b}), we need to introduce some further
notation.  We define,
\begin{equation}
  \label{eq:redpm}
\ud{\widehat{\Gamma}_{(u,\mathrm{dec})\pm.i}}{ab}{a'b'C}  = 
\ud{\widehat{\Gamma}_{(u,\mathrm{dec})0.i}}{ab}{a'b'C} \pm
\sqrt{s}\,\ud{\widehat{\Gamma}_{(u,\mathrm{dec})1.i}}{ab}{a'b'C}~,
\qquad
i=0,\ldots,3~,
\end{equation}
and
\begin{equation}
  \label{eq:N}
  \N = \left.\sqrt{p^0+\mb_a}\sqrt{p^{'0}+\mb_{a'}}\right|_{CMF}
     = \frac{1}{2\sqrt{s}}\sqrt{(\sqrt{s}-\mb_a)^2-\mm_b^2}
       \sqrt{(\sqrt{s}-\mb_{a'})^2-\mm_{b'}^2}~. 
\end{equation}
Notice that $\N$ depends on flavor indices through baryon and meson
masses.  Its dependence, like that of $\uav$ and $\uwd$ defined in
appendix \ref{sec:appkin}, is not made explicit in the notation .  The
reduced partial waves of (\ref{eq:dec-u-pwvs}) are then, omitting
flavor indices for simplicity,
\begin{subequations}
    \label{eq:red-dec-u-pwvs-all}
  \begin{equation}
    \label{eq:red-dec-u-pwvs}
    \widehat{f}_{(u,\mathrm{dec})\ell +} = H_\ell - \frac{1}{\ell+1}
    K_\ell~,
    \qquad
    \widehat{f}_{(u,\mathrm{dec})\ell -} = H_\ell + \frac{1}{\ell} K_\ell~,   
  \end{equation}
with
\begin{equation}
  \label{eq:red-dec-u-pwvs-aux}
    H_\ell = \N \sum_{i=0}^3 \widehat{\Gamma}_{(u,\mathrm{dec})+.i}
    h^{(\ell)}_{+.i} - \frac{\Delta u}{\N} \sum_{i=0}^3
    \widehat{\Gamma}_{(u,\mathrm{dec})-.i} h^{(\ell)}_{-.i}~,
\qquad
    K_\ell = - \frac{\Delta u}{\N} \sum_{i=0}^3
    \widehat{\Gamma}_{(u,\mathrm{dec})-.i} k^{(\ell)}_{-.i}~.    
\end{equation}
\end{subequations}
The integrals $h^{(\ell)}_{\pm.i}$, $k^{(\ell)}_{-.i}$, which like
$\widehat{\Gamma}_{(u,\mathrm{dec})\pm.i}$ depend on $a$,
$b$, $a'$, $b'$, $C$, are defined in appendix \ref{sec:ints}.

\subsection{Partial waves from nonet-baryon exchange}
\label{sec:pwnnt}

We consider the exchange of nonet baryon resonances with $J^P=3/2^-$,
whose mostly-singlet member $\Lambda(1520)$ features prominently in
$pK^-$ scattering cross sections.  We describe these resonances as a
nonet of Rarita-Schwinger fields, $\N_0$, $\N=1/\sqrt{2} \sum_{a=1}^8
\N_a \beta^a$, with the same flavor-matrix representation for $\N$ as
for the ground-state baryon octet \cite{bou08}.  These fields are
taken to be mass eigenstates, with $\N_0=\Lambda(1520)$,
$\N_{1,2,3}=\Sigma(1620)^{+,-,0}$, $\N_{4,6}=N(1520)^{+,0}$,
$\N_{5,7}=\Xi(1820)^{-,0}$, $\N_8=\Lambda(1690)$.  Thus, their
interaction Lagrangian must take into account the mixing of singlet
and octet components of $\N_{0,8}$, but their tree-level propagator is
diagonal.  For each $\N_a$, its free propagator has the same form as
for each member of the decuplet (see, e.g., eq.\ (3.8) of
\cite{mei00}).

The interaction vertices for nonet and octet baryons and pseudoscalar
mesons are described by a Lagrangian with the same Lorentz structure
as the Lagrangian for decuplet baryons (except for an additional
$\gamma_5$ due to the opposite intrinsic parity of nonet and
decuplet), and with the same flavor structure as the Lagrangian for
ground-state baryons and mesons, augmented by $\N_0$--$\N_8$ mixing.
Explicitly
\begin{equation}
  \label{eq:nonlag}  
\begin{aligned}
  f\mathcal{L}_{\mathrm{nnt}} &= \sum_{h=1}^7 \udu{\F_{(8)}}{c}{d}{h}
  \partial^\mu\phi_c \overline{B}^d (g_{\mu\nu}-\kappa_9
  \gamma_\mu\gamma_\nu)\gamma_5 \N_h^\nu 
  +\cos\theta  \udu{\F_{(8)}}{c}{d}{8}
  \partial^\mu\phi_c \overline{B}^d (g_{\mu\nu}-\kappa_9
  \gamma_\mu\gamma_\nu)\gamma_5 \N_8^\nu \\
  &+\sin\theta  \udu{\F_{(8)}}{c}{d}{8}
  \partial^\mu\phi_c \overline{B}^d (g_{\mu\nu}-\kappa_9
  \gamma_\mu\gamma_\nu)\gamma_5 \N_0^\nu 
  +\cos\theta D_0 \tr(\partial^\mu\phi_c \overline{B}^d)
  (g_{\mu\nu}-\kappa_9 \gamma_\mu\gamma_\nu)\gamma_5 \N_0^\nu \\  
  &-\sin\theta D_0 \tr(\partial^\mu\phi_c \overline{B}^d)
  (g_{\mu\nu}-\kappa_9 \gamma_\mu\gamma_\nu)\gamma_5 \N_8^\nu  +
  \mathrm{h.c.}~,\\
  \udu{\F_{(8)}}{c}{d}{h} &\equiv D_8 \udu{d}{c}{d}{h} - F_8 \udu{f}{c}{d}{h}~,
\end{aligned}
\end{equation}
which establishes the definition of our coupling constants and mixing
parameter.  In (\ref{eq:nonlag}) $D_0$, $D_8$ and $F_8$ are the
singlet, and the $D$- and $F$-type octet, couplings, resp., and $\theta$
is the nonet mixing angle.  It is also apparent in
$\mathcal{L}_{\mathrm{nnt}}$ that we are using the same off-shell
parameter $\kappa_9$ for both singlet and octet fields \cite{lut00}.
While there is to our knowledge no reason why those parameters should
be precisely equal, we have found that keeping the singlet and octet
parameters independent does not lead to improvements in the
description of experimental data.  Actually, in our experience, the
best fits are obtained by setting them to be equal.

The tree-level amplitudes obtained from (\ref{eq:nonlag}) can be
parameterized as,
\begin{equation}
  \label{eq:nntamp}
  \begin{aligned}
    \ud{\T_\mathrm{nnt}}{ab}{a'b'} &= \ud{\T_{(s,\mathrm{nnt})}}{ab}{a'b'} +
    \ud{\T_{(u,\mathrm{nnt})}}{ab}{a'b'}~,\\ 
    \ud{\T_{(s,\mathrm{nnt})}}{ab}{a'b'} &= \frac{1}{f^2}\sum_{e=0}^8
    \ud{\F_{(s,\mathrm{nnt})}}{abe}{a'b'} \frac{1}{s-\Mbnt_e^2+i0}
    \overline{u}'
    \left(\ud{\widehat{\Gamma}_{(s,\mathrm{nnt})0}}{abe}{a'b'} +
      \ud{\widehat{\Gamma}_{(s,\mathrm{nnt})1}}{abe}{a'b'}\ptirac\right) u~,\\ 
    \ud{\T_{(u,\mathrm{nnt})}}{ab}{a'b'} &= \frac{1}{f^2}\sum_{e=0}^8
    \ud{\F_{(u,\mathrm{nnt})}}{abe}{a'b'} \frac{1}{u-\Mbnt_e^2+i0}
    \overline{u}'
    \left(\ud{\widehat{\Gamma}_{(u,\mathrm{nnt})0}}{abe}{a'b'} +
      \ud{\widehat{\Gamma}_{(u,\mathrm{nnt})1}}{abe}{a'b'}\ptirac\right) u~.
  \end{aligned}
\end{equation}
The flavor index $e$ in the sum runs through the octet
($e=1,\ldots,8$) and singlet ($e=0$) baryon resonances, with $\Mbnt_e$
the mass of the $e^\mathrm{th}$ nonet member.  The flavor
coefficients in (\ref{eq:nntamp}) are,
\begin{equation}
  \label{eq:nntflv}
  \begin{aligned}
    \ud{\F_{(s,\mathrm{nnt})}}{abe}{a'b'} &=
    \left(D_8\udd{d}{e}{b'}{a'}-F_8\udd{f}{e}{b'}{a'}\right)
    \left(D_8\uud{d}{b}{a}{e}+F_8\uud{f}{b}{a}{e}\right),
    \qquad e=1,\ldots,7~,\\
    \ud{\F_{(s,\mathrm{nnt})}}{ab8}{a'b'} &= \left(\cos\theta
      \left(D_8\udd{d}{8}{b'}{a'}-F_8\udd{f}{8}{b'}{a'}\right)
      -\sin\theta D_0 e_{b'a'} \right)\!
    \left(\cos\theta
      \left(D_8\uud{d}{b}{a}{8}+F_8\uud{f}{b}{a}{8}\right)
      -\sin\theta D_0 e^{ba} \right),\\
    \ud{\F_{(s,\mathrm{nnt})}}{ab0}{a'b'} &= \left(\sin\theta
      \left(D_8\udd{d}{8}{b'}{a'}-F_8\udd{f}{8}{b'}{a'}\right)
      +\cos\theta D_0 e_{b'a'} \right)\!
    \left(\sin\theta
      \left(D_8\uud{d}{b}{a}{8}+F_8\uud{f}{b}{a}{8}\right)
      +\cos\theta D_0 e^{ba} \right),\\
    \ud{\F_{(u,\mathrm{nnt})}}{abe}{a'b'} &=
    \left(D_8\uud{d}{b}{e}{a'}+F_8\uud{f}{b}{e}{a'}\right)
    \left(D_8\uud{d}{a}{b'}{e}-F_8\uud{f}{a}{b'}{e}\right),
    \qquad e=1,\ldots,7~,\\
    \ud{\F_{(u,\mathrm{nnt})}}{ab8}{a'b'} &= \left(\cos\theta
      \left(D_8\uud{d}{b}{8}{a'}+F_8\uud{f}{b}{8}{a'}\right)
      -\sin\theta D_0 e^{b}_{a'} \right)\!
    \left(\cos\theta
      \left(D_8\udd{d}{a}{b'}{8} - F_8\udd{f}{a}{b'}{8}\right)
      -\sin\theta D_0 e^{a}_{b'} \right),\\
    \ud{\F_{(u,\mathrm{nnt})}}{ab0}{a'b'} &= \left(\sin\theta
      \left(D_8\uud{d}{b}{8}{a'}+F_8\uud{f}{b}{8}{a'}\right)
      +\cos\theta D_0 e^{b}_{a'} \right)\!
    \left(\sin\theta
      \left(D_8\udd{d}{a}{b'}{8} - F_8\udd{f}{a}{b'}{8}\right)
      +\cos\theta D_0 e^{a}_{b'} \right)~.\\
  \end{aligned}  
\end{equation}
The reduced amplitudes $\widehat{\Gamma}_{(s,\mathrm{nnt})0,1}$ and
$\widehat{\Gamma}_{(u,\mathrm{nnt})0,1}$ in (\ref{eq:nntamp}) are
obtained from those for the decuplet by substituting $\Mb_C
\rightarrow -\Mbnt_e$ and $\kappa \rightarrow \kappa_9$.

From (\ref{eq:nntamp}) the partial waves for nonet-exchange amplitudes
are seen to take the form,
\begin{equation}
  \label{eq:nntpw}
  \begin{aligned}
        \ud{f_{(s,\mathrm{nnt})\ell\pm}}{ab}{a'b'} &= \frac{1}{f^2}
        \sum_{e=0}^8 \ud{\F_{(s,\mathrm{nnt})}}{abe}{a'b'}
        \frac{1}{s-\Mbnt_e^2+i0}
        \ud{\widehat{f}_{(s,\mathrm{nnt})\ell\pm}}{abe}{a'b'}~,\\
         \ud{f_{(u,\mathrm{nnt})\ell\pm}}{ab}{a'b'} &= \frac{1}{f^2}
        \sum_{e=0}^8 \ud{\F_{(u,\mathrm{nnt})}}{abe}{a'b'}
        \ud{\widehat{f}_{(u,\mathrm{nnt})\ell\pm}}{abe}{a'b'}~.\\
  \end{aligned}
\end{equation}
In the $s$-channel case, the partial waves with $j=3/2$ take a
particularly simple form,
\begin{equation}
  \label{eq:nntpws}
  \begin{aligned}
       \ud{f_{(s,\mathrm{nnt})1+}}{ab}{a'b'} &= \frac{1}{f^2}
       \left(-\frac{\N\uwd}{6}\right)
        \sum_{e=0}^8 \ud{\F_{(s,\mathrm{nnt})}}{abe}{a'b'}
        \frac{1}{\sqrt{s}+\Mbnt_e}~,\\
       \ud{f_{(s,\mathrm{nnt})2-}}{ab}{a'b'} &= \frac{1}{f^2}
       \left(-\frac{\uwd^2}{12\N}\right)
        \sum_{e=0}^8 \ud{\F_{(s,\mathrm{nnt})}}{abe}{a'b'}
        \frac{1}{\sqrt{s}-\Mbnt_e}~,\\
  \end{aligned}
\end{equation}
where $\N$ is defined in (\ref{eq:N}) and $\uwd$ in (\ref{eq:kinnot}).
Since, as discussed in section \ref{sec:res}, only the $D_{3/2}$
partial wave for nonet exchange enters our analysis, we omit the
explicit expressions of the $S_{1/2}$ and $P_{1/2}$ waves for brevity.
For the $u$-channel, the reduced partial waves in (\ref{eq:nntpw}) can
be obtained from those for decuplet-exchange,
(\ref{eq:red-dec-u-pwvs-all}), by means of the substitutions $\Mb_C
\rightarrow -\Mbnt_e$ and $\kappa \rightarrow \kappa_9$ in the reduced
amplitudes, and the substitution $\Mb_C \rightarrow \Mbnt_e$ in the
integrals $h^{(\ell)}_{\pm.i}$, $k^{(\ell)}_{-.i}$ given in appendix
\ref{sec:ints}.

\subsection{Partial waves from vector-meson exchange}
\label{sec:pwvec}

We expect vector-meson interactions to play a significant role in the
energy range $0\leq q_\mathrm{lab}\lesssim 600$ MeV considered in this
paper.  In tree-level chiral perturbation theory, vector mesons
contribute to meson--baryon scattering through $t$-channel exchange.
The Lagrangians for pseudoscalar meson--vector meson and ground-state
baryon--vector meson interactions have been discussed in
\cite{mei00,eck89,bor96} and refs.\ therein.  Here we follow the
notation of \cite{mei00}.  The coupling constant for the PPV meson
vertex is denoted $G_V$ (expressed here in MeV), and there are eight
leading-order couplings for baryon--vector meson interactions, denoted
$R_{D,F}$, $S_{D,F}$, $T_{D,F}$, $U_{D,F}$, \cite{mei00} with units
MeV$^{0,-1,-3,-2}$ resp.  In the $t$-channel vector-meson exchange
amplitude these couplings appear as products $G_V X_{D,F}$, with
$X=R,S,T,U$.  The value of $G_V\sim 60$ MeV can be estimated from
vector meson decays, or from pseudoscalar-meson electromagnetic form
factors \cite{mei00}.  The values of the coupling constants in the
three-flavor baryon--vector meson Lagrangian are not well established.

The tree-level $t$-channel vector-meson exchange contribution to the
$T$ matrix can be parameterized as,
\begin{equation}
  \label{eq:vecamp}
  \begin{aligned}
  \ud{\T_{(\mathrm{vec})}}{ab}{a'b'} &= \frac{4\sqrt{2}G_V}{f^2}\sum_{X=R,S,T,U}\sum_{c=1}^8
  \ud{f}{cb}{b'}\dud{\X}{c}{a}{a'} \frac{1}{t-\Mv_c^2+i0}
  \overline{u}'
  \left(\ud{\widehat{\Gamma}_{(X)0}}{abc}{a'b'} +
    \ud{\widehat{\Gamma}_{(X)1}}{abc}{a'b'}\ptirac\right) u~,\\
  \dud{\X}{c}{a}{a'} &= \dud{d}{c}{a}{a'}X_D + \dud{f}{c}{a}{a'}X_F~,
  \qquad
  X = R,S,T,U~,
  \end{aligned}
\end{equation}
where $\X=\R,\S,\T,\U$ for $X=R,S,T,U$, resp.  We expand the reduced
amplitudes in (\ref{eq:vecamp}) as,
\begin{equation}
  \label{eq:redvecamp}
  \ud{\widehat{\Gamma}_{(X)0}}{abc}{a'b'} =
  \ud{\widehat{\Gamma}_{(X)0.0}}{abc}{a'b'} +
  \ud{\widehat{\Gamma}_{(X)0.1}}{abc}{a'b'} t +
  \ud{\widehat{\Gamma}_{(X)0.2}}{abc}{a'b'} t^2 +
  \ud{\widehat{\Gamma}_{(X)0.3}}{abc}{a'b'} u~,
\end{equation}
and analogously $\ud{\widehat{\Gamma}_{(X)1}}{abc}{a'b'}$.  The
coefficients in the r.h.s.\ of (\ref{eq:redvecamp}) are independent of
$u$ and $t$.  Computation at tree level yields the non-vanishing
coefficients,
  \begin{equation}
    \label{eq:vecampcoef}
    \begin{gathered}
  \ud{\widehat{\Gamma}_{(R)0.0}}{abc}{a'b'} =-\frac{1}{2}
  \left(s+(\mb_a+\mb_{a'})^2\right)~,
\;\;\;
  \ud{\widehat{\Gamma}_{(R)0.3}}{abc}{a'b'} =\frac{1}{2}~,
\;\;\;
\ud{\widehat{\Gamma}_{(R)1.0}}{abc}{a'b'} =\mb_a+\mb_{a'}~,\\
  \ud{\widehat{\Gamma}_{(S)0.0}}{abc}{a'b'} =\frac{1}{4}
  (\mb_{a'}-\mb_a)(\mm_{b'}^2-\mm_b^2)~,
\;\;\;
  \ud{\widehat{\Gamma}_{(S)0.1}}{abc}{a'b'} =
  -\frac{1}{4}(\mb_a+\mb_{a'})~, 
\;\;\;
\ud{\widehat{\Gamma}_{(S)1.1}}{abc}{a'b'} =\frac{1}{2}~,\\
  \ud{\widehat{\Gamma}_{(T)0.0}}{abc}{a'b'} =\frac{1}{16}
  (\mb_{a'}-\mb_a)^2
  (\mb_a+\mb_{a'})(2s+2\mb_{a'}\mb_a-\mm_b^2-\mm_{b'}^2)~, \\
  \ud{\widehat{\Gamma}_{(T)0.1}}{abc}{a'b'} = \frac{1}{16}
  (\mb_{a}-\mb_{a'})(\mb_a^2-\mb_{a'}^2+\mm_{b'}^2-\mb_b^2)~,\\
  \ud{\widehat{\Gamma}_{(T)0.2}}{abc}{a'b'} =\frac{1}{16}
  (\mb_a+\mb_{a'})~,
\;\;\; 
  \ud{\widehat{\Gamma}_{(T)1.0}}{abc}{a'b'} =-\frac{1}{8}
  (\mb_{a'}^2-\mb_a^2)^2~, 
\;\;\;
  \ud{\widehat{\Gamma}_{(T)1.2}}{abc}{a'b'} =-\frac{1}{8}\\
  \ud{\widehat{\Gamma}_{(U)0.0}}{abc}{a'b'} =\frac{1}{8}
  (\mb_{a}-\mb_{a'}) \left(\mb_{a}\mb_{a'}(\mb_{a}-\mb_{a'}) -
    \mb_{a}\mm_{b}^2 + \mb_{a'} \mm_{b'}^2\right) + \frac{1}{8}
  (\mb_{a}-\mb_{a'})^2 s~,\\
  \ud{\widehat{\Gamma}_{(U)0.1}}{abc}{a'b'}
  =\frac{\mb_{a}^2+\mb_{a'}^2}{8},
\;\;\;
  \ud{\widehat{\Gamma}_{(U)1.0}}{abc}{a'b'} = -\frac{\mb_{a}+\mb_{a'}}{8} 
  (\mb_{a}-\mb_{a'})^2~,
\;\;\;
  \ud{\widehat{\Gamma}_{(U)1.1}}{abc}{a'b'}
  = -\frac{\mb_{a}+\mb_{a'}}{8}.
    \end{gathered}
  \end{equation}
As above, we need to introduce the linear combinations
\begin{equation}
  \label{eq:redvecpm}
\ud{\widehat{\Gamma}_{(\mathrm{X})\pm.i}}{ab}{a'b'c}  = 
\ud{\widehat{\Gamma}_{(\mathrm{X})0.i}}{ab}{a'b'c} \pm
\sqrt{s}\,\ud{\widehat{\Gamma}_{(\mathrm{X})1.i}}{ab}{a'b'c}~,
\qquad
i=0,\ldots,3,~~X=R,S,T,U.
\end{equation}
The partial waves then have the form,
\begin{equation}
  \label{eq:vecpw}
  \ud{f_{(\mathrm{vec})\ell\pm}}{ab}{a'b'} = \frac{4\sqrt{2}G_V}{f^2}\sum_{X=R,S,T,U}\sum_{c=1}^8
  \ud{f}{cb}{b'}\dud{\X}{c}{a}{a'} \ud{\widehat{f}_{(X)\ell\pm}}{abc}{a'b'}~,
\end{equation}
with the reduced partial waves of (\ref{eq:redvecpm}) given by,
omitting flavor indices for simplicity,
\begin{subequations}
    \label{eq:red-vec-all}
  \begin{equation}
    \label{eq:red-vec-pwvs}
    \widehat{f}_{(X)\ell +} = H_{(X)\ell} - \frac{1}{\ell+1}
    K_{(X)\ell}~,
    \qquad
    \widehat{f}_{(X)\ell -} = H_{(X)\ell} +
    \frac{1}{\ell} K_{(X)\ell}~,    
  \end{equation}
with
\begin{equation}
  \label{eq:red-vec-pwvs-aux}
    H_{(X)\ell} = \N \sum_{i=0}^3 \widehat{\Gamma}_{(X)+.i}
    \widetilde{h}^{(\ell)}_{+.i} - \frac{\Delta t}{\N} \sum_{i=0}^3
    \widehat{\Gamma}_{(X)-.i} \widetilde{h}^{(\ell)}_{-.i}~,
\qquad
    K_{(X)\ell} = - \frac{\Delta t}{\N} \sum_{i=0}^3
    \widehat{\Gamma}_{(X)-.i} \widetilde{k}^{(\ell)}_{-.i}~.    
\end{equation}
\end{subequations}
The parameterization (\ref{eq:redvecamp}) was chosen so that the
integrals $\widetilde{h}^{(\ell)}_{\pm.i}$,
$\widetilde{k}^{(\ell)}_{-.i}$ appearing in
(\ref{eq:red-vec-pwvs-aux}) can be obtained from those in
(\ref{eq:red-dec-u-pwvs-aux}) by means of a substitution, as indicated
in appendix \ref{sec:ints}.

\subsection{Partial waves from $\mathcal{O}(q^2)$ contact interactions}
\label{sec:q2pw}

At low energies the meson--baryon contact vertices provide the largest
contributions to the scattering amplitude.  We incorporate in our
Lagrangian flavor-symmetry breaking $\mathcal{O}(q^2)$ corrections to
contact interactions.  The $\mathcal{O}(q^2)$ Lagrangian \cite{oll06a}
contains fifteen terms\footnote{But only seven independent
  combinations contribute to the meson--baryon scattering amplitude
  \cite{oll06z}.} out of which only three are flavor-breaking. 
The corresponding coupling constants are conventionally denoted
$b_{0,D,F}$ \cite{oll06a}. The the tree-level meson--baryon scattering
amplitude from these contact vertices are,
\begin{equation}
  \label{eq:ampq2}
  \begin{aligned}
  \ud{\T_{(q^2)}}{ab}{a'b'} &= -\frac{1}{f^2}
  \ud{\F_{(q^2)}}{ab}{a'b'} \overline{u}'u~,\\
  \ud{\F_{(q^2)}}{ab}{a'b'}&= (3b_0+2b_D) e^a_{a'}
  (\chi_0e^b_{b'}+\chi_8 \uud{d}{8}{b}{b'}) 
  + \frac{2}{3}\chi_8 b_D (e^8_{b'} \uud{d}{b}{a}{a'} + e^{8b}
  \dud{d}{b'}{a}{a'} + e^b_{b'} \uud{d}{8}{a}{a'}) \\
  &+ \frac{2}{3}\chi_8 b_F (e^8_{b'} \uud{f}{b}{a}{a'} + e^{8b}
  \dud{f}{b'}{a}{a'} + e^b_{b'} \uud{d}{8}{a}{a'}) 
  + \sum_r 3\chi_0 \udd{d}{b}{b'}{r} (b_D \uud{d}{r}{a}{a'} + b_F
  \uud{f}{r}{a}{a'})\\
  &+\sum_{r,s} \chi_8 (b_D \uud{d}{r}{a}{a'} + b_F \uud{f}{r}{a}{a'})
  (\udd{d}{b}{b'}{s} \uud{d}{8}{s}{r} + \udd{d}{8}{b'}{s}
  \uud{d}{s}{b}{r} + \udd{d}{8}{b}{s} \uud{d}{s}{b'}{r})~.
  \end{aligned}
\end{equation}
Flavor violations in the $\mathcal{O}(q^2)$ meson Lagrangian
\cite{gas84} are described by the matrix $\chi = 2B_0 \mathcal{M}_q$,
with $\mathcal{M}_q$ the quark mass matrix and $B_0 = -\langle 0|
q\overline{q} |0 \rangle/f^2$ the quark condensate.  In the
isospin-symmetry limit $\mathcal{M}_q = \diag(\hat{m},\hat{m},m_s)$
and we can parameterize $\chi$ as,
\begin{equation}
  \label{eq:chi}
  \chi = \chi_0 I + \chi_8 \beta^8~.
\end{equation}
This equation defines the parameters $\chi_{0,8}$ appearing in
(\ref{eq:ampq2}).  At $\mathcal{O}(q^2)$ and in the isospin limit the
meson \cite{gas84} and baryon \cite{oll06a} Lagrangians  lead to the
relations,
\begin{equation}
  \label{eq:masses}
  \begin{aligned}
 \widetilde{m}_b^2 &= \chi_0 + \chi_8 \uud{d}{8}{b}{b}~, \\
 m_a &= M_\mathrm{aux} - 4 b_D\chi_8 \uud{d}{8}{a}{a} -
 4b_F\chi_8 \uud{f}{8}{a}{a}~,
 \qquad
 M_\mathrm{aux} \equiv M_0 - 6b_0\chi_0- 4b_D\chi_0~,    
  \end{aligned}
\end{equation}
with $M_0$ on the second line being the common mass of the
ground-state baryon octet in the chiral limit.  These relations will
be used below to fix, or estimate, the values of $\chi_{0,8}$ and
$b_{0,D,F}$.

The amplitudes (\ref{eq:ampq2}) lead to the partial waves,
\begin{equation}
  \label{eq:q2pw}
  \ud{f_{(q^2)0}}{ab}{a'b'} = -\frac{1}{f^2} \N \ud{\F_{(q^2)}}{ab}{a'b'}~,
  \qquad
  \ud{f_{(q^2)1-}}{ab}{a'b'} = \frac{1}{f^2} \frac{\uwd}{2\N} \ud{\F_{(q^2)}}{ab}{a'b'}~,
\end{equation}
with $\N$ defined in (\ref{eq:N}) and $\uwd$ in appendix
\ref{sec:appkin}.  All other partial waves vanish.

\section{Unitarized partial waves}
\label{sec:untr}

Tree-level BChPT is not sufficient by itself to describe three-flavor
meson-baryon dynamics in the $S=-1$ sector, in which strong coupling
effects such as subthreshold resonances render BChPT inapplicable even
at threshold.  We unitarize the tree-level amplitudes with the method
of \cite{mei00,oll01}.  A technically detailed explanation of the
method can be found in those references.  We shall limit ourselves
here to stating the result of the unitarization of tree-level
amplitudes.

Given a set of coupled reaction channels $|B^{a_i}M^{b_i}\rangle
\rightarrow |B^{a_j}M^{b_j}\rangle$, we denote $(f_{\ell\pm})_{ij} =
\pw{\ell\pm}{a_i b_j}{a_j b_j}$ the corresponding tree-level
partial-wave matrix.  A solution to the unitarity equation for $T$,
resumming the right-hand cut in the $s$-plane, is given by the partial
waves $(\mathcal{F}_{\ell\pm})_{ij}$ related to $(f_{\ell\pm})_{ij}$
by the matrix equation,
\begin{equation}
  \label{eq:unitarization}
  \mathcal{F}_{\ell\pm} = \left( I + f_{\ell\pm} \cdot G \right)^{-1}
  \cdot f_{\ell\pm}~,
\end{equation}
where $I$ is an identity matrix, and $G$ is the diagonal matrix
$G_{ij} = g^{a_i b_i} \delta_{ij}$ (no summation over $i$, $j$).  The
``unitarity bubbles'' $g^{ab}$ are given by,
\begin{equation}\label{eq:bubble}
  \begin{split}
  g^{ab}(s) &= \frac{i\mu^\epsilon}{(2\pi)^d} \int d^dk \frac{1}
  {(k^2-\mb^2+i 0)((k+p_T)^2 - m_a^2+i 0)}
  \quad +~~ \mathrm{counterterm}\\
  &= \frac{1}{16\pi^2} \left\{ a^{ab} +
      \log\left(\frac{\mb_a^2}{\mu^2}\right) +
     \frac{s-\mb_a^2+\mm_b^2}{2s} \log\left( \frac{\mm_b^2}{\mb_a^2}
     \right) + \frac{w(s,\mb_a^2,\mm_b^2)}{2s} \right. \times \\
  &\times \left\{\left[  \log\left( \frac{\mb_a^2 - \mm_b^2 -s -
          w(s,\mb_a^2,\mm_b^2)}{2s} -i 0  \right) -
      \log\left( \frac{\mb_a^2 - \mm_b^2 -s +
          w(s,\mb_a^2,\mm_b^2)}{2s} + i 0  \right)
        \right]
    \rule{0pt}{16pt}\right\}~,
  \end{split}
\end{equation}
with $w(x,y,z)$ defined in appendix \ref{sec:appkin}.  The loop function
$g^{ab}$ was computed in (\ref{eq:bubble}) in dimensional
regularization.  The subtraction constants $a^{ab}$, depending on the
renormalization scale $\mu$, are taken as free parameters in each
isospin channel.  Variations in $\mu$ can be offset by a redefinition
of $a^{ab}$ \cite{mei00}. 

\section{Results}
\label{sec:res}

In this section we discuss our results for the reactions $p K^-
\rightarrow B^{a'} M^{b'}$.  Within the range of initial-meson
momentum $0\leq q_\mathrm{lab}\lesssim 600$ MeV considered here the
only possible final states are $N\overline{K}$, $\Lambda\pi$,
$\Sigma\pi$.  Following \cite{ose98,oll01,jid02,ben02}, however, we
apply the unitarization method of \cite{mei00,oll01} including
as intermediate states also $\Lambda\eta$, $\Sigma\eta$, $\Xi K$.  This
is justified by the fact that in lowest-order BChPT those states are
degenerate.  The contribution of $D$ waves is obviously important in
the region around and above the $\Lambda(1520)$ resonance, but
$F$ waves are negligible for $q_\mathrm{lab}\lesssim 800$ MeV.
Thus, we compute physical observables from $S_{1/2}$, $P_{1/2}$,
$P_{3/2}$, $D_{3/2}$ and $D_{5/2}$ partial waves obtained from
ten-channel unitarization of tree-level partial waves.  

The baryon nonet requires separate consideration.  We find that it is
not possible, as pointed out in \cite{lut00}, to fit experimental data
if partial waves from nonet-baryon exchange other than $D_{3/2}$ are
included.  This is undoubtedly a reflection of nonet resonances being
dynamical in nature, or at least possessing a significant dynamical
component.  Unlike \cite{lut00}, however, we find that by including
both the $s$ and $u$ channels we get somewhat lower $\chi^2$ values
than by dropping the $u$-channel contribution.  We therefore take into
account only the $D_{3/2}$ wave in nonet-baryon exchange amplitudes,
but retain both $s$- and $u$-channel contributions to it. As a
consequence, the parameter $\kappa_9$ does play a role in our fits.

The experimental data included in our fits consists of about 2800
points comprising the threshold branching fractions $\gamma$, $R_c$
and $R_n$ \cite{tov71,nov78}, total
\cite{sak65,arm70,mas75,mas76,ban81,cib82,eva83} and differential
\cite{arm70,mas76} cross section data for the six open reactions
channels up to $q_\mathrm{lab}\lesssim 600$ MeV, and the first two
Legendre moments of the CM-frame differential cross sections and spin
asymmetries for $pK^-\rightarrow \Sigma\pi, \Lambda\pi$
\cite{mas75,ban81,cib82}.  In the figures below we display also
higher-energy cross-section data from those references and from
\cite{ada75,als78}, and CM-frame spin asymmetry data from
\cite{arm70}, which were not included in our fits.

\subsection{Fitting procedure and parameters}
\label{sec:fitting}

For numerical computations we set meson and baryon masses to their
physical values \cite{PDG}.  Following \cite{oll01,jid02}, below we
set $\mu=630$ MeV.  The coupling constants for the ground-state baryon
and meson octet vertices have been computed from semileptonic hyperon
decays \cite{bor99,rat99}.  We keep them fixed at the values $D=0.80$
and $F=0.46$, which are consistent with \cite{rat99} and with the
tree-level results of \cite{bor99}.  The coupling constants for the
nonet of baryon resonances can be obtained from a flavor SU(3)
analysis of their tree-level strong decay widths \cite{pla70,lut00}.
Since, however, in our treatment those resonances acquire their widths
dynamically through the loop corrections involved in unitarization, we
take their couplings as free parameters in our fits.

In order to obtain a picture of the role played by the different
interactions described in sect.\ \ref{sec:treel}, we present below
results from three different fits to data.  First, we consider
scattering amplitudes obtained from the ground-state
baryon--pseudoscalar meson $\mathcal{O}(q^1)$ contact interaction, and
from $s$- and $u$-channel exchange of octet, decuplet and nonet
baryons.  The best fit obtained with those amplitudes is referred to
below as ``fit III.''  Second, a different series of fits is obtained
by augmenting the previous amplitudes by $t$-channel vector-meson
exchange.  For these fits we take the coupling-constant products $G_V
X_{D,F}$, in the notation of sect.\ \ref{sec:pwvec}, as free
parameters.  The best of those fits is denoted ``fit II'' below.   

Finally, we report on another series of fits obtained by adding to the
amplitudes of fit II the $\mathcal{O}(q^2)$ flavor-symmetry breaking
corrections to baryon--meson contact vertices.  The LECs in the
$\mathcal{O}(q^2)$ amplitudes of sect.\ \ref{sec:q2pw} are related to
pseudoscalar meson masses and ground-state baryon mass splittings.
Fitting the expression for meson masses (\ref{eq:masses}) to data
\cite{PDG} we obtain
\begin{equation}
\label{eq:mesmass}
  \chi_0 = (412.04\;\mathrm{ MeV})^2~, \qquad \chi_8 = -(510.54\;
  \mathrm{MeV})^2~.
\end{equation}
We keep $\chi_{0,8}$ fixed to these values in our fits, which actually
implies no restriction since in the flavor-breaking
$\mathcal{O}(q^2)$ contact interactions, couplings always appear in
the combinations $\chi_i b_j$ with $i=0,8$, $j=0,D,F$.  The couplings
$b_{0,D,F}$ can then be either treated as free parameters in our
partial waves or, alternatively, fixed from baryon mass data. Fitting 
the baryon masses in (\ref{eq:masses}) to experimental data \cite{PDG}
with $M_\mathrm{aux}$, $b_{D,F}$ as parameters we get
\begin{equation}
  \label{eq:barmass}
  M_0 - 6 \chi_0 b_0 = 1110.61\;\mathrm{MeV}~,
  \qquad
  b_D = 6.51\times 10^{-5}\;\mathrm{MeV}^{-1}~,
  \qquad
  b_F = -2.17\times 10^{-4}\;\mathrm{MeV}^{-1}~.
\end{equation}
This equation does not completely fix $b_0$ due to the uncertainty in
$M_0$, with $-3\times 10^{-4} < b_0 < -2\times 10^{-4}$ MeV$^{-1}$ for
$800<M_0<900$ MeV.  We call ``fit I'' the best fit obtained by setting
$b_{D,F}$ to the values (\ref{eq:barmass}), and taking $b_0$ as a free
parameter constrained to the range (-3)---(-2) $\times 10^{-4}$
MeV$^{-1}$.  As pointed out in \cite{oll06z}, however, because
unitarization resums an infinite sequence of diagrams, the numerical
values for $b_{0,D,F}$ we obtain in UBChPT need not be the same as
those obtained in fixed-order BChPT.  Thus, we also performed a fit
with freely varying $b_{0,D,F}$, referred to as ``fit I$'$''.  As
discussed below, the numerical values for $b_{0,D,F}$ obtained in fit
I$'$ are close to those of I, and the plots of fits I and I$'$ are
virtually identical.

A summary of our best-fit parameters is as follows.
\begin{trivlist}
\item[\textbf{Fit I}] $f=90.82$ MeV, $b_0 = -3.0\times 10^{-4}$ MeV$^{-1}$,
  \begin{trivlist}
  \item Subtraction constants:
\begin{equation*}
  a^{N\overline{K}} = -1.83, ~
  a^{\Lambda\pi} = -2.14,    ~
  a^{\Sigma\pi} = -0.63,     ~
  a^{\Lambda\eta} = -1.75,   ~
  a^{\Sigma\eta} = -2.14,    ~
  a^{\Xi K} =  -1.41~.
\end{equation*}
  \item Decuplet and nonet parameters: 
\begin{equation*}
g_{10}=0.89,~
\kappa=-0.32,~
D_0 =1.68,~
D_8 =0.20,~
F_8 =1.59,~
\kappa_9 =2.49,~
\theta =-0.33~.  
\end{equation*}
\item Vector-meson couplings: $[R_{D,F}]=1$,
  $[S_{D,F}]=\mathrm{MeV}^{-1}$, $[T_{D,F}]=\mathrm{MeV}^{-3}$,
  $[U_{D,F}]=\mathrm{MeV}^{-2}$, 
\begin{gather*}
  G_V R_D = 52.76  ,~
  G_V R_F = 75.42  ,~
  G_V S_D = 0.16  ,~
  G_V S_F = 0.036 ,~
\\
  G_V T_D = 1.23\times 10^{-6}  ,~
  G_V T_F = -1.36\times 10^{-7}   ,~
  G_V U_D = 1.27\times 10^{-4}  ,~
  G_V U_F = 1.37\times 10^{-4} ~ .
\end{gather*}
  \end{trivlist}
\item[\textbf{Fit I$'$}] $f=91.42$ MeV,
  \begin{trivlist}
  \item Subtraction constants:
\begin{equation*}
  a^{N\overline{K}} = -1.84, ~
  a^{\Lambda\pi} = -1.72,    ~
  a^{\Sigma\pi} = -0.70,     ~
  a^{\Lambda\eta} = -1.70,   ~
  a^{\Sigma\eta} = -2.15,    ~
  a^{\Xi K} =  -0.78~.
\end{equation*}
  \item Decuplet and nonet parameters: 
\begin{equation*}
g_{10}=0.90,~
\kappa=-0.31,~
D_0 =1.65,~
D_8 =-0.03,~
F_8 =1.89,~
\kappa_9 =2.31,~
\theta =-0.34~.  
\end{equation*}
\item Vector-meson couplings: $[R_{D,F}]=1$,
  $[S_{D,F}]=\mathrm{MeV}^{-1}$, $[T_{D,F}]=\mathrm{MeV}^{-3}$,
  $[U_{D,F}]=\mathrm{MeV}^{-2}$, 
\begin{gather*}
  G_V R_D = 52.87  ,~
  G_V R_F = 68.75  ,~
  G_V S_D = 0.19  ,~
  G_V S_F = 0.022 ,~
\\
  G_V T_D = 9.83\times 10^{-7}  ,~
  G_V T_F = -4.96\times 10^{-8}   ,~
  G_V U_D = 1.46\times 10^{-4}  ,~
  G_V U_F = 2.00\times 10^{-4} ~ .
\end{gather*}
  \item $\mathcal{O}(q^2)$ contact interaction couplings: [MeV$^{-1}$]
\begin{equation*}
  b_0=-4.0\times 10^{-4},\quad b_D=1.45\times 10^{-4},\quad b_F=-2.38\times 10^{-4}~.   
\end{equation*}
  \end{trivlist}
\item[\textbf{Fit II}] $f=95.97$ MeV,
  \begin{trivlist}
  \item Subtraction constants:
\begin{equation*}
  a^{N\overline{K}} =  -1.84, ~
  a^{\Lambda\pi} =  -3.31,    ~
  a^{\Sigma\pi} =   -0.41,     ~
  a^{\Lambda\eta} = -3.28,   ~
  a^{\Sigma\eta} =  -2.54,    ~
  a^{\Xi K} =     -1.15~.
\end{equation*}
\item Decuplet and nonet parameters:
\begin{equation*}
g_{10}=0.87,~
\kappa=-0.34,~
D_0 =1.68,~
D_8 =-0.038,~
F_8 =1.84,~ 
\kappa_9 =2.28,~
\theta =-0.31~.  
\end{equation*}
\item Vector-meson couplings:  $[R_{D,F}]=1$,
  $[S_{D,F}]=\mathrm{MeV}^{-1}$, $[T_{D,F}]=\mathrm{MeV}^{-3}$,
  $[U_{D,F}]=\mathrm{MeV}^{-2}$, 
\begin{gather*}
  G_V R_D = 57.90 ,~
  G_V R_F = 89.91 ,~
  G_V S_D = 3.66\times 10^{-3} ,~
  G_V S_F = -0.12 ,~
\\
  G_V T_D = 5.27\times 10^{-7}  ,~
  G_V T_F = -5.0\times 10^{-7}  ,~
  G_V U_D =-1.70\times 10^{-5}  ,~
  G_V U_F =1.20\times 10^{-4} ~ .
\end{gather*}
  \end{trivlist}
\item[\textbf{Fit III}] $f=95.81$ MeV,
  \begin{trivlist}
  \item Subtraction constants:
\begin{equation*}
  a^{N\overline{K}}  = -1.68, ~
  a^{\Lambda\pi} =  -2.16,    ~
  a^{\Sigma\pi} =  -0.82 ,     ~
  a^{\Lambda\eta} = -3.85,   ~
  a^{\Sigma\eta} =  -2.56,    ~
  a^{\Xi K} =     -1.27~.
\end{equation*}
\item Decuplet and nonet parameters:
\begin{equation*}
g_{10}=0.83,~
\kappa=-0.28,~
D_0 = 1.73 ,~  
D_8 = 0.20  ,~  
F_8 = 1.74  ,~  
\kappa_9 = 2.30  ,~  
\theta= -0.17~.
\end{equation*}
  \end{trivlist}
\end{trivlist}
Some remarks about these parameters are in order.  The subtraction
constants show less dispersion about their ``natural'' dimensional
regularization value -2 \cite{oll01} in fits I and I$'$ than in fit
II, which shows less dispersion than III.  Most of the parameters are
quite stable across fits, with the exception of $D_8$, $\theta$ and
$G_V S_{D,F}$ which show larger variations.  Other determinations of
$b_{0,D,F}$ from fits to $pK^-$ data are given in \cite{lut00,oll06z},
with results quite similar to ours.

The values for the nonet couplings $D_0$, $D_8$, $F_8$ are similar to
those expected from the flavor symmetry analysis of decay widths of
\cite{pla70}, as updated in \cite{lut00}: 1.57, 0.59, 1.27, resp.,
with $D_8$ showing the largest departures.  The values for $\theta$ we
obtain are in general agreement with the observation that
$\Lambda(1520)$ is predominantly a flavor singlet.  For fits I, I$'$
and II we get $\theta\simeq -20^o$ to be compared with the value
$-28^o$ adopted in \cite{lut00}.  As remarked above, the nonet
parameters reported here differ from those computed in
\cite{pla70,lut00} by loop corrections, so numerical equality among
them is not expected.

A direct comparison of our numerical results for vector-meson
couplings in three-flavor UBChPT to those of the two-flavor analysis
of pion--nucleon scattering of \cite{mei00} would not be meaningful.
It is nevertheless interesting to find out where our results stand
relative to those of \cite{mei00}.  Assuming $G_V=60$ MeV, from
\cite{mei00} we find,
\begin{equation}
  \label{eq:mei00}
  \begin{gathered}
G_V(R_D+R_F)\simeq 312\;\mathrm{MeV}~,
\quad    
G_V (S_D+S_F -\frac{m_N}{2}(U_D+U_F)) \simeq 3.3\times 10^{-1}~, \\
6\times 10^{-7} \lesssim G_V(T_D+T_F) \lesssim 1.8\times 10^{-6}
\;\mathrm{MeV}^{-2} ~. 
  \end{gathered}
\end{equation}
From our fit I we get $G_V(R_D+R_F) = 128.2$ MeV,
$G_V(T_D+T_F)=1.1\times 10^{-6}$ MeV$^{-2}$, which are of the same
order of magnitude as (\ref{eq:mei00}), and $G_V (S_D+S_F
-\frac{m_N}{2}(U_D+U_F))=2.1\times 10^{-2}$ which is an order of
magnitude smaller than the value in (\ref{eq:mei00}).  This latter
result, however, arises from a numerical cancellation between the two
terms and is therefore rather fortuitous.  Indeed, we could fine-tune
the couplings $U_{D,F}$ in our fit so as to lead to a larger value
roughly in agreement with (\ref{eq:mei00}) without appreciably
changing the fit results.  As is easy to check, fit I$'$ leads to the
same conclusions. From fit II we get, $G_V(R_D+R_F) = 147.8$ MeV,
$G_V(T_D+T_F)=2.7\times 10^{-8}$ MeV$^{-2}$ and $G_V (S_D+S_F
-\frac{m_N}{2}(U_D+U_F))=1.7\times 10^{-1}$.  Whereas in all our
type-I fits we find $-T_F\ll T_D$, in our type-II fits we
systematically obtain $-T_F\sim T_D$, causing $G_V(T_D+T_F)$ to be
small compared to (\ref{eq:mei00}).

\subsection{Physics results}
\label{sec:physres}

For fit I we get the threshold branching fractions (defined in App.\
\ref{sec:obsv}) $\gamma$ = 2.35 (2.36$\pm$ 0.04), $R_c$ = 0.645
(0.664$\pm$0.011), $R_n$ = 0.210 (0.189$\pm$ 0.015), in good agreement
with the experimental values \cite{tov71,nov78} quoted in parentheses.
Fits I$'$, II and III lead to essentially the same results.  We
computed also the scattering length $a_{K^-p}$ (see App.\
\ref{sec:obsv}) for the elastic process at threshold, which is not
included in our fits.  From fit I we get $a_{K^-p} = -1.09 + 0.63 i$
fm, with the other fits showing only variations in the second decimal.
This value for $a_{K^-p}$ is in reasonable agreement with the KEK
result \cite{kek1,kek2}, $a_{K^-p} = (-0.78\pm 0.15\pm 0.03)+ i
(0.49\pm 0.25\pm 0.12)$ fm, but significantly larger than the DEAR one
\cite{bee05}, $a_{K^-p} = (-0.468\pm 0.09\pm 0.015)+ i (0.302\pm
0.135\pm 0.036)$ fm.  The issue of the (in)consistency of the DEAR
measurement with the previous KEK one and with other hadronic data is
beyond the scope of this paper; it has been discussed in detail in
\cite{mei04,bor05x,oll05x,bor05y,oll06z} and references cited there.

\begin{figure}[h]
  \centering
\includegraphics{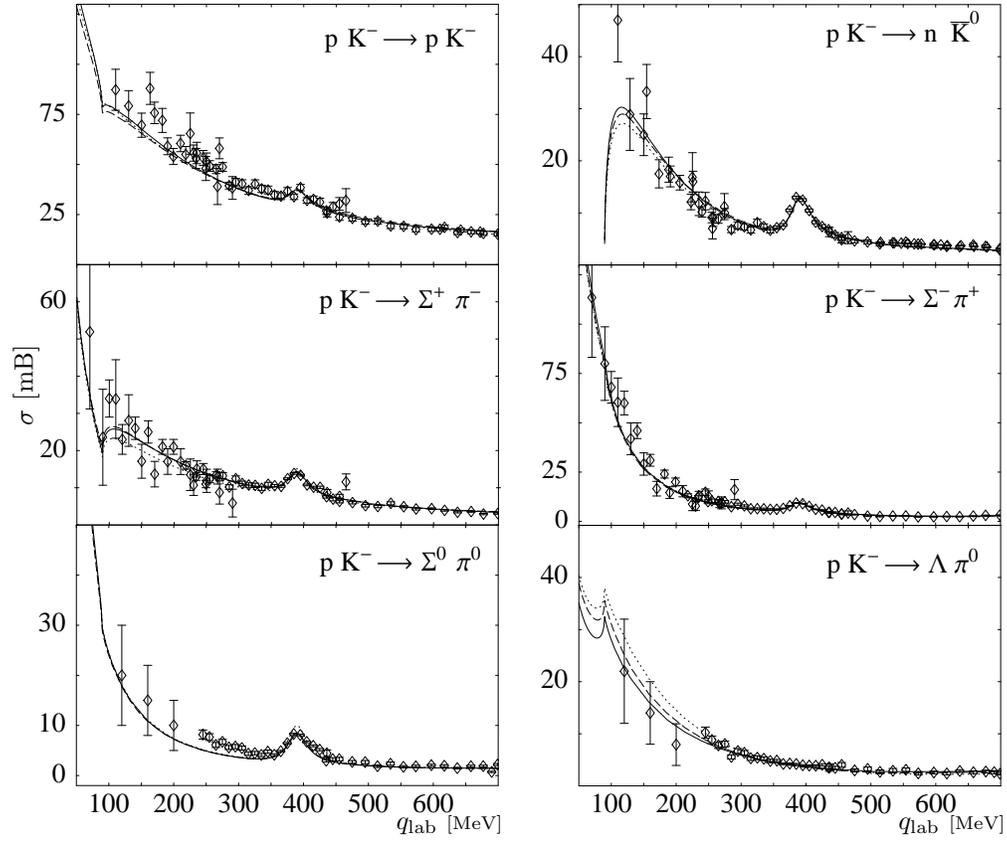}  
\caption{Total cross sections for $pK^-$ scattering. Solid lines: fit
  I, which includes all amplitudes described in sect.\
  \ref{sec:treel}.  Dashed lines: fit II, which includes only
  $\mathcal{O}(q^1)$ amplitudes. Dotted lines: fit III, which includes
  $\mathcal{O}(q^1)$ amplitudes except vector-meson exchange ones.  Data
  from \cite{sak65,arm70,mas75,mas76,ban81,cib82,eva83,ada75,als78}.}
  \label{fig:1}
\end{figure}

We present our results for cross sections and spin asymmetries in the
figures below.  For clarity, we omit plotting fit I$'$ in the figures,
since its curves are almost indistinguishable from fit I.  Fig.\
\ref{fig:1} shows our results for total cross sections.  Fits I 
and II give a very good description of data, while fit III is somewhat
less accurate, especially in the region of the $\Lambda(1520)$ peak
where it tends to overshoot the data in $\Sigma\pi$ channels.  All
three fits, and a large number of other fits we have conducted,
underestimate the $\Sigma^0\pi^0$ cross section for $q_\mathrm{lab} <
350$ MeV. We cannot explain this phenomenon, which is also present to
different extents in \cite{lut00,jid02,bou08}. 

We included all data for differential cross sections from
\cite{arm70,mas76} up to $q_\mathrm{lab}=600$ MeV in our fits, though
some higher energy data is also shown fig.\ \ref{fig:2}.  Given the
large number of data points reported in \cite{arm70,mas76}, only a
representative selection of results is shown in the figure. The
overall agreement of fits I and II with data is excellent, while fit
III provides a less accurate though still reasonably good description.
For all three fits the agreement with data is better for
charged-baryon final states than for neutral ones. As seen in the
figure, for charged-baryon channels even data with $q_\mathrm{lab}$
over 600 MeV is well reproduced by fits I and II.

In fig.\ \ref{fig:3} we show our results for Legendre moments of
differential cross sections and of CMF spin asymmetries normalized to
$A_0$ (see App.\ \ref{sec:obsv} for definitions).  We omit the moments
for processes with nucleon final states because very detailed data on
differential cross sections for them has already been included in our
fits.  For the processes with hyperon final states shown in the figure
we restrict ourselves to the first two moments, $A_{1,2}$ and
$B_{1,2}$, since higher-order ones have rather small values and large
experimental errors which make them consistent with zero throughout
the energy range. The experimental data on those moments are
displayed in the figure without modification.  Thus, to match the
experimentally measured quantities, the moments $B_{1,2}$ computed
with our unitarized amplitudes are multiplied by the final-state
baryon polarizabilities $\alpha$ reported in \cite{PDG}.
As seen in fig.\ \ref{fig:3}, the data for most processes have small
errors in the region of the $\Lambda(1520)$ resonance, providing a
good constraint to fit parameters.  The agreement with data is
globally very good, especially for the charged modes
$pK^-\rightarrow\Sigma^\pm\pi^\mp$. Fits I and II accurately reproduce
the structures in the data around the $\Lambda(1520)$ peak.  For the
moment $B_1$ in $pK^-\rightarrow\Lambda\pi^0$ all three fits yield
positive values at lower energies, with a negative slope, whereas as
remarked in \cite{bou08}, both \cite{lut00, bou08} obtain negative
values for that moment.

Fig.\ \ref{fig:4} shows our results for the CMF final-state spin
asymmetry.  The asymmetry data were not included in our fits due to
their rather large experimental errors.  In that sense, these results
are ``predictions'' of the theory.  Interestingly, for this observable
the three fits show some of the largest differences among them.  Thus,
more precise data on spin asymmetry, polarization or analyzing power
could provide some of the most discriminating and stringent tests of
the theory.  As seen in the figure, within experimental errors all
three fits describe the data very well at all energies.  Since these
data were not fitted, such agreement is non-trivial.

Finally, in fig.\ \ref{fig:5} we display the $\Sigma\pi$ mass
distribution computed from the isoscalar components of the amplitudes
for $N\overline{K} \rightarrow \Sigma\pi$ and $\Sigma\pi \rightarrow
\Sigma\pi$ in $S_{1/2}$ wave, and for $N\overline{K} \rightarrow
\Sigma\pi$ in $D_{3/2}$ wave.  From the $S_{1/2}$ wave we obtain the
mass spectrum for the resonance $\Lambda(1405)$, dynamically generated
by unitarization \cite{ose98,lut00,oll01,jid02}, which is well known
to be process dependent.  That dependence can in principle be
understood from the two-pole structure of this resonance \cite{jid03}.
The spectra plotted in the figure are consistent with those of
\cite{jid03} (see fig.\ 4 of that ref.  See also, e.g., figs.\ 9 of
\cite{lut00} and 2 of \cite{ose98}).  Averaging the results from the
three fits, we obtain a resonance peak at $1409.7\pm 1.5$ MeV with a
width of $20.2\pm 0.5$ MeV for the $\Sigma\pi \rightarrow \Sigma\pi$
channel, and a peak at $1414.7\pm 1$ MeV with a width of $20\pm 1$ MeV
for the $N\overline{K} \rightarrow \Sigma\pi$ channel.  Similarly,
from the $D_{3/2}$ wave we obtain the mass spectrum for
$\Lambda(1520)$.  Again averaging the results from the three fits, we
obtain a resonance peak at $1518.5\pm 0.5$ MeV and a width of $14.1\pm
0.8$ MeV, which are in excellent agreement with the measured values
\cite{PDG} if we take into account that three-body decay modes not
considered here constitute 11\% of the full experimental width.

In summary, an excellent global description of $pK^-$ scattering data
is obtained with fits I and II, up to $q_{\mathrm{lab}}\lesssim 600$
MeV, and including threshold observables.  Fit III is in very good
agreement with data up to $q_{\mathrm{lab}}\lesssim 350$ MeV, but
provides a weaker description of data in the region of the
$\Lambda(1520)$ peak and above.  Vector-meson interactions play an
important role in the theoretical description, leading to a decrease
of $\sim 25\%$ in the global $\chi^2$ of fit II with respect to III,
and to an improved description of the data in the $\Lambda(1520)$
region as can be seen in the figures.  Inclusion of $\mathcal{O}(q^2)$
flavor-breaking contact interactions leads to some additional, though
less marked, improvements reflected in a further decrease of $\sim
5\%$ in the global $\chi^2$.  Experimental results can be
theoretically reproduced only if nonet baryon resonance exchange is
restricted to the $D_{3/2}$ wave, which is strong evidence that those
resonances are produced by the coupled-channel dynamics or, at least,
possess a large dynamical component.  Whereas the global agreement
with data is remarkably good, there are differences in detail between
theory and data, especially in reaction channels with a neutral final
baryon, and in most channels at momenta $q_{\mathrm{lab}}\gtrsim 600$.
We attribute those differences mostly to the contribution of processes
with two final mesons, particularly $pK^-\rightarrow
\Sigma^0\pi^0\pi^0$, which are not included in our treatment.

\section{Final remarks}
\label{sec:finrem}

In the previous sections we presented a detailed study of two-body
polarized $pK^-$ scattering in the energy range from threshold through
the $\Lambda(1520)$ peak, $0\leq q_\mathrm{lab}\lesssim 600$ MeV, in
UBChPT.  Our results show excellent global agreement with experimental
data.  This is achieved by taking into account ground-state baryon and
meson interactions, including flavor-breaking contact interactions of
$\mathcal{O}(q^2)$, as well as $J^P=3/2^+$ decuplet and $3/2^-$ nonet
exchange in $s$- and $u$-channels, and $t$-channel vector-meson
exchange.  Notice that the latter has not been considered in previous
treatments of $pK^-$ scattering in UBChPT.  Whereas scattering
amplitudes involving solely pseudoscalar mesons yield good
semiquantitative agreement with data, inclusion of vector-meson
exchange diagrams leads to an improved quantitative description.
Further small improvements are also obtained from the flavor-breaking
contact interactions.  The five partial waves $S_{1/2}$, $P_{1/2}$,
$P_{3/2}$, $D_{3/2}$, $D_{5/2}$ were taken into account in all
amplitudes, with the exception of baryon nonet exchange ones.  The
fact that it is necessary to exclude all waves but $D_{3/2}$ of
nonet-mediated diagrams originates most certainly in the large
dynamical component of $J^P=3/2^-$ resonances reported in the
literature.

From the theoretical point of view, we report explicit expressions for
partial waves for $s$- and $u$-channel baryon, and for $t$-channel
vector-meson, exchange diagrams not given in the previous literature.
From a phenomenological point of view, an improvement with respect to
previous global analyses of $pK^-$ low-energy data is our inclusion of
differential cross-section data in the region of the $\Lambda(1520)$
resonance and above.  Our fits describe the data remarkably well,
including measured threshold parameters, total and differential
cross-section data, and spin asymmetries.  A noteworthy example is the
description of the CMF spin asymmetry (in the form of Legendre
moments, see fig.\ \ref{fig:3}) around the $\Lambda(1520)$ peak, which
also shows the important role played by vector meson interactions.
Very good agreement is also obtained with the spin asymmetry data
shown in fig.\ \ref{fig:4}, although those data were not included in
the fit.  As seen in figs. \ref{fig:3} and \ref{fig:4}, spin
observables are quite discriminating among different fits.  More
precise spin-asymmetry data would be theoretically most desirable and
challenging.  For the fitted couplings and, especially, subtractions
constants we obtain more natural values than in our previous
lower-energy treatment \cite{bou08}.  A further improvement in the
analysis, and an extension in its energy range, should be made
possible by the addition of further reaction channels to the
unitarization procedure, particularly three-body processes.  Work
along those lines is currently in progress.

\section*{Acknowledgements}

The author gratefully acknowledges access to the computer
``KanBalam'' granted to him by Departamento de Superc\'omputo,
Direcci\'on General de Servicios de C\'omputo Acad\'emico, Universidad
Nacional Aut\'onoma de M\'exico.

\newpage
\appendix

\renewcommand{\theequation}{\thesection.\arabic{equation}}
\setcounter{equation}{0}
\section{Kinematics}
\label{sec:appkin}

In this appendix we gather some kinematical definitions used 
throughout the paper.  We introduce the notation
\begin{equation}
  \label{eq:omega}
  \omega(x,y,z) = (x^2+y^2+z^2-2xy-2xz-2yz)^\frac{1}{2}
  = (x-(\sqrt{y}+\sqrt{z})^2)^\frac{1}{2}
  (x-(\sqrt{y}-\sqrt{z})^2)^\frac{1}{2}~. 
\end{equation}
The function $\omega$ appears frequently in relativistic kinematics
(\emph{e.g.,} in the center of mass frame $|\vec{p}| =
\omega(s,\mb_a^2, \mm_{b}^2)/(2\sqrt{s})$). The Mandelstam invariants
for the process $| B^{a}(p,\sigma) M^b(q)\rangle \longrightarrow |
B^{a'}(p',\sigma') M^{b'}(q')\rangle$ are
\begin{equation}
  \label{eq:mandelstam}
  s=(p+q)^2=(p'+q')^2~,
  \quad
  t=(p-p')^2=(q-q')^2~,
  \quad
  u=(p-q')^2=(p'-q)^2~,
\end{equation}
with $s+t+u = \mb_a^2 + \mb_{a'}^2 + \mm_{b}^2 + \mm_{b'}^2~$.
The physical region for the process is defined by the inequalities
\begin{equation}
  \label{eq:physreg}
  s_\sss{\mathrm{th}}\leq s~,
  \quad
  \tmin \leq t \leq \tmax~,  
  \quad
  \umin \leq u \leq \umax~,
\end{equation}
where,
\begin{equation}
  \label{eq:kinlim}
  \begin{aligned}
    s_\sss{\mathrm{th}} &= \max\left\{(\mb_a + \mm_b)^2, (\mb_{a'} +
      \mm_{b'})^2 \right\}\\
    t_{\substack{\sss{\mathrm{max}}\\[-2pt]\sss{\mathrm{min}}}} &=
    -\frac{1}{2s} \left( \rule{0pt}{10pt} s^2 - s (\mb_a^2 +
      \mb_{a'}^2 + \mm_b^2 + \mm_{b'}^2 ) + (\mb_a^2 - \mm_b^2)
      (\mb_{a'}^2 - \mm_{b'}^2) \right) \\
    &\quad \pm \frac{1}{2s} \omega(s,\mb_a^2,\mm_b^2)
      \omega(s,\mb_{a'}^2,\mm_{b'}^2),\\
    u_{\substack{\sss{\mathrm{max}}\\[-2pt]\sss{\mathrm{min}}}} &=
    -\frac{1}{2s} \left( \rule{0pt}{10pt} s^2 - s (\mb_a^2 +
      \mb_{a'}^2 + \mm_b^2 + \mm_{b'}^2 ) - (\mb_a^2 - \mm_b^2)
      (\mb_{a'}^2 - \mm_{b'}^2) \right) \\
    &\quad \pm \frac{1}{2s}
    \omega(s,\mb_a^2,\mm_b^2) \omega(s,\mb_{a'}^2,\mm_{b'}^2).\\    
  \end{aligned}
\end{equation}
We introduce also the following useful notations,
\begin{equation}
  \label{eq:kinnot}
\uav = \frac{\umax+\umin}{2}~,
\quad
\tav = \frac{\tmax+\tmin}{2}~,
\quad
\uwd = \frac{\umax-\umin}{2}=\frac{\tmax-\tmin}{2}=\twd~.
\end{equation}
As is easy to check, $s+\tav+\uav = \mb_a^2 +
\mb_{a'}^2 + \mm_{b}^2 + \mm_{b'}^2~$ and, in the CMF,
\begin{equation}
  \label{eq:kinnot2}
  u = \uav - x\uwd~, \quad t = \tav + x\twd~, \quad x\equiv
  \widehat{p}'\cdot \widehat{p}~.
\end{equation}
In the laboratory frame we have, in terms of Mandelstam invariants,
\begin{subequations}
  \label{eq:labfrm}
\begin{equation}
  \label{eq:labfrma}
  q^{0}_\mathrm{lab} = \frac{1}{2\mb_a}(s - \mb_a^2 - \mm_{b}^2),
\quad
  q'^{0}_\mathrm{lab} = \frac{1}{2\mb_a}(\mb_a^2+\mm_{b'}^2-u),
\quad
  p'^{0}_\mathrm{lab} = \frac{1}{2\mb_a}(\mb_a^2+\mb_{a'}^2-t),
\end{equation}
therefore,
\begin{equation}
  \label{eq:labfrmb}
q_\mathrm{lab}\equiv |\vec{q}_\mathrm{lab}| = \frac{1}{2\mb_a}
\omega(s,\mb_a^2,\mm_b^2), 
\quad 
|\vec{q}\,'_\mathrm{lab}| = \frac{1}{2\mb_a}
\omega(u,\mb_a^2,\mm_{b'}^2), 
\quad
|\vec{p}\,'_\mathrm{lab}| = \frac{1}{2\mb_a} \omega(t,\mb_a^2,\mb_{a'}^2). 
\end{equation}
\end{subequations}

\setcounter{equation}{0}
\section{Partial-wave integrals}
\label{sec:ints}

The integrals of Legendre polynomials $P_\ell(x)$, and their
derivatives, involved in the expression of partial waves in
(\ref{eq:red-dec-u-pwvs-all}) and (\ref{eq:red-vec-all})
 are listed in this appendix.  For decuplet baryons
$\Mb_C$ denotes the mass of the $C^\mathrm{th}$ member of the
decuplet, $C=1,\ldots,10$.  We denote also,
\begin{equation}
  \label{eq:zC}
  z_C = \frac{\Mb_C^2-\uav}{\uwd}~,
\end{equation}
with $\uav$, $\uwd$ defined in appendix \ref{sec:appkin}.  In what
follows, $Q_\ell (z)$ denotes the Legendre function of the second kind
\cite{abram}, analytic on the $z$ plane cut along $-1< z < 1$ for
$\ell$ a nonnegative integer.  When analytic continuation is necessary
(e.g., if $-1<z_C<1$ in the expressions below) the mass $\Mb_C$ should
be understood as $\Mb_C - i 0$. The invariants $u$ and $t$ are given
as functions of $x=\cos\theta_{CM}$ in (\ref{eq:kinnot2}).
\begin{align}
  \label{eq:ints}
  h^{(\ell)}_{+.0} &= \int_{-1}^{1}\!dx\,\frac{1}{u-M_C^2}P_\ell(x)
  = \frac{(-1)^{\ell+1}}{\uwd}Q_\ell(z_C)\\
  h^{(\ell)}_{+.1} &= \int_{-1}^{1}\!dx\,\frac{u}{u-M_C^2}P_\ell(x)
  = (-1)^{\ell+1} \left(\frac{M_C^2}{\uwd}Q_\ell(z_C)-\delta_{\ell 0}\right)\\
  h^{(\ell)}_{+.2} &= \int_{-1}^{1}\!dx\,\frac{u^2}{u-M_C^2}P_\ell(x)
  = (-1)^{\ell+1} \left(\frac{(\uav+ z_C\uwd)^2}{\uwd}Q_\ell(z_C) -
    (2\uav+z_C\uwd)\delta_{\ell 0} - \frac{\uwd}{3} \delta_{\ell 1}
  \right)\\
  h^{(\ell)}_{+.3} &= \int_{-1}^{1}\!dx\,\frac{t}{u-M_C^2}P_\ell(x) =
  (-1)^{\ell+1} \left\{\left(\frac{\tav}{\uwd}-z_C\right) Q_\ell(z_C)
    + \delta_{\ell 0}\right\} \\
  h^{(\ell)}_{-.0} &= \int_{-1}^{1}\!dx\,\frac{x}{u-M_C^2}P_\ell(x) =
\frac{(-1)^{\ell}}{\uwd}(z_CQ_\ell(z_C)-\delta_{\ell 0})\\
h^{(\ell)}_{-.1} &= \int_{-1}^{1}\!dx\,\frac{x u}{u-M_C^2}P_\ell(x) =
(-1)^{\ell} \left\{ z_C \left(z_C+\frac{\uav}{\uwd}\right) Q_\ell(z_C) -
  \left(z_C+\frac{\uav}{\uwd}\right)\delta_{\ell 0} -
  \frac{1}{3}\delta_{\ell 1} \right\} \\
h^{(\ell)}_{-.2} &= \int_{-1}^{1}\!dx\,\frac{x u^2}{u-M_C^2}P_\ell(x) =
(-1)^{\ell} \left\{ \rule{0pt}{15pt}
  \frac{z_C}{\uwd} \left(\uav + z_C\uwd\right)^2 Q_\ell(z_C) \right.\nonumber\\
 &\left. -\frac{1}{\uwd} \left((\uav+z_C\uwd)^2+\frac{1}{3}\uwd^2 \right)
\delta_{\ell 0}  
-\frac{1}{3}(2\uav+z_C\uwd) \delta_{\ell 1} - \frac{2}{15}\uwd
\delta_{\ell 2} \rule{0pt}{15pt}\right\}\\
h^{(\ell)}_{-.3} &= \int_{-1}^{1}\!dx\,\frac{x t}{u-M_C^2}P_\ell(x) =
(-1)^{\ell+1} \left\{ z_C\left(z_C-\frac{\tav}{\uwd}\right) Q_\ell(z_C)
- \left(z_C-\frac{\tav}{\uwd}\right) \delta_{\ell 0} - \frac{1}{3}
\delta_{\ell 1} \right\} \\
  k^{(\ell)}_{-.0} &= \int_{-1}^{1}\!dx\,\frac{1-x^2}{u-M_C^2}P'_\ell(x) =
\frac{(-1)^{\ell+1}}{\uwd}
\frac{\ell(\ell+1)}{2\ell+1}(Q_{\ell+1}(z_C)-Q_{\ell-1}(z_C)) \\
  k^{(\ell)}_{-.1} &= \int_{-1}^{1}\!dx\,\frac{(1-x^2)u}{u-M_C^2}P'_\ell(x) =
  (-1)^{\ell+1} \frac{\ell(\ell+1)}{2\ell+1}
  \left\{ \left(z_C +
      \frac{\uav}{\uwd}\right)(Q_{\ell+1}(z_C)-Q_{\ell-1}(z_C))
    +\delta_{\ell 1} \right\}\\ 
  k^{(\ell)}_{-.2} &=
  \int_{-1}^{1}\!dx\,\frac{(1-x^2)u^2}{u-M_C^2}P'_\ell(x) =
  (-1)^{\ell+1} \frac{\ell(\ell+1)}{2\ell+1} \left\{ \frac{1}{\uwd}
    (\uav + z_C\uwd)^2
    (Q_{\ell+1}(z_C)-Q_{\ell-1}(z_C))\right. \nonumber\\
  &-\left.  \frac{\uwd}{3} \delta_{\ell 0} + (2\uav +
  z_C\uwd)\delta_{\ell 1} + \frac{\uwd}{3} \delta_{\ell 2} \right\} \\
  k^{(\ell)}_{-.3} &=
  \int_{-1}^{1}\!dx\,\frac{(1-x^2)t}{u-M_C^2}P'_\ell(x) =
  (-1)^{\ell+1} \frac{\ell(\ell+1)}{2\ell+1} \left\{
    \left(-z_C+\frac{\tav}{\uwd}\right)
    (Q_{\ell+1}(z_C)-Q_{\ell-1}(z_C)) - \delta_{\ell 1} \right\}
\end{align}
For the vector-meson exchange partial waves in (\ref{eq:red-vec-all})
we define,
\begin{equation}
  \label{eq:yc}
   y_c = \frac{\Mv_c^2 - \tav}{\twd}~,\qquad c=1,\ldots,8~.
\end{equation}
Then, the integrals in (\ref{eq:red-vec-pwvs-aux}) can be obtained
from the ones above with the substitutions.  
\begin{equation}
  \label{eq:vecint}
  \widetilde{h}^{(\ell)}_{\pm.i} =
  \left[ \rule{0pt}{20pt}h^{(\ell)}_{\pm.i}\right]\!\!{}_{\substack{\uav\rightarrow\tav \\
      \uwd\rightarrow -\twd \\ z_C\rightarrow -y_c}}~,
\qquad
  \widetilde{k}^{(\ell)}_{\pm.i} =
  \left[ \rule{0pt}{20pt}k^{(\ell)}_{\pm.i}\right]\!\!{}_{\substack{\uav\rightarrow\tav \\
      \uwd\rightarrow -\twd \\ z_C\rightarrow -y_c}}~.
\end{equation}
Notice that $Q_\ell(-z) = (-1)^{\ell+1} Q_\ell(z)$.

\setcounter{equation}{0}
\section{Physics observables}
\label{sec:obsv}

In this appendix we summarize the expressions in terms of partial
waves of the physics observables considered in sect.\ \ref{sec:res}. 
The amplitude for the process $| B^{a}(p,\sigma) M^b(q)\rangle
\longrightarrow | B^{a'}(p',\sigma') M^{b'}(q')\rangle$ is
parameterized as,
\begin{equation}
  \label{eq:paramT}
  \ud{\T}{ab}{a'b'}\equiv \langle B_{a'}(p',\sigma') M_{b'}(q') | T |
  B^{a}(p,\sigma) M^b(q) \rangle = \overline{u}'
  (\ud{\Gamma_0}{ab}{a'b'} + \ud{\Gamma_1}{ab}{a'b'} 
  \ptirac) u~.
\end{equation}
The associated partial-wave expansion is given in (\ref{eq:pw}).
Below, we omit flavor indices for simplicity.
The total and differential cross sections are given in terms of
partial waves by,
\begin{align}
  \label{eq:crsct}
\sigma &= \frac{(\hbar c)^2}{32\pi s}
\frac{\omega(s,\mb_{a'}^2,\mm_{b'}^2)}{\omega(s,\mb_{a}^2,\mm_{b}^2)}
\sum_{\ell=0}^\infty \frac{2}{2\ell+1} \left(|(\ell+1) f_{\ell+}+\ell
  f_{\ell-}|^2 + \ell (\ell+1) |f_{\ell+} - f_{\ell-}|^2\right),\\
\frac{d\sigma}{d\Omega} &= \frac{(\hbar c)^2}{64\pi^2 s}
\frac{\omega(s,\mb_{a'}^2,\mm_{b'}^2)}{\omega(s,\mb_{a}^2,\mm_{b}^2)}
\left( \left|\sum_{\ell=0}^\infty \left((\ell+1) f_{\ell+} + \ell
      f_{\ell-} \right) P_\ell(x) \right|^2 +
  (1-x^2) \left|\sum_{\ell=0}^\infty \left( f_{\ell+} -
    f_{\ell-} \right) P_\ell(x) \right|^2 \right),
\end{align}
with the function $\omega$ defined in (\ref{eq:omega}) and $x = 
  \widehat{p}'\cdot \widehat{p}$ in the CMF.

If we denote $\P'$ the polarization of the final baryon in the
CMF, the CMF spin asymmetry is given by \cite{bou08a}, 
\begin{equation}
  \label{eq:CMasym}
  \begin{aligned}
  \A'\equiv \frac{d\sigma}{d\Omega}\P' &= \frac{(\hbar c)^2}{64\pi^2 s}
\frac{\omega(s,\mb_{a'}^2,\mm_{b'}^2)}{\omega(s,\mb_{a}^2,\mm_{b}^2)} 
\Im(\Gamma_0^* \Gamma_1) \left(-i \Tr{\pirac \ptirac \pirac' \not\!s'
  \gamma_5} \right) ~,\\
  &= \frac{(\hbar c)^2}{64\pi^2 s^{3/2}}
\omega(s,\mb_{a'}^2,\mm_{b'}^2)^2 \sqrt{1-x^2}
\Im(\Gamma_0^* \Gamma_1) ~,
  \end{aligned}
\end{equation}
where on the second line we set $s'^\mu=(0,\widehat{s}')$ with
$\widehat{s}' = (\widehat{p}\wedge \widehat{p}')/|\widehat{p}\wedge
\widehat{p}'|$, since we are interested in polarization orthogonal to
the reaction plane, and used the definition of CM frame.  Expanding in
partial waves up to $\ell=2$ we obtain,
\begin{equation}
  \label{eq:CMasymfinal}
  \begin{aligned}
  \A' &= \frac{(\hbar c)^2}{32\pi^2 s} 
\frac{\omega(s,\mb_{a'}^2,\mm_{b'}^2)}{\omega(s,\mb_{a}^2,\mm_{b}^2)} 
\sqrt{1-x^2} \left\{
\Im\left((f_{1+}^* - f_{1-}^*)f_{0}\right) +
\Im\left((f_{1+}^*-f_{1-}^*)(3f_{2+} + 2f_{2-})\right) P_2(x) 
\right. \\
& + \Im\left((f_{2+}^*-f_{2-}^*)f_{0}\right) P_2'(x) +
\Im\left((f_{2+}^* - f_{2-}^*)(2f_{1+} + f_{1-})\right) P_2'(x) P_1(x)
+ 3\Im\left(f_{1+}^*f_{1-}\right) P_1(x) \\
&+ \left. 5
\Im\left(f_{2+}^*f_{2-}\right) P_2'(x) P_2(x) \right\}~.
  \end{aligned}
\end{equation}
The Legendre moments of the differential cross section and spin
asymmetry are defined as \cite{mas75,ban81},
\begin{equation}
  \label{eq:legmom}
\frac{d\sigma}{d\Omega} = \sum_{n=0}^\infty A_n P_n(x)~,
\qquad
\frac{d\sigma}{d\Omega}\P' = \sqrt{1-x^2} \sum_{n=1}^\infty B_n
P_n'(x)~.
\end{equation}
The moments $A_n$, $B_n$ can be expressed in terms of partial
waves.  For notational convenience we define,
\begin{equation}
  \label{eq:auxa}
  a_0 \equiv |f_{0}|^2 + \frac{1}{3} |2f_{1+} + f_{1-}|^2 +
  \frac{1}{5} |3f_{2+} + 2f_{2-}|^2 + \frac{2}{3} |f_{1+}-f_{1-}|^2 +
  \frac{6}{5} |f_{2+}-f_{2-}|^2~.
\end{equation}
We thus have the partial-wave expansions, up to $D$ wave,
\begin{equation}
  \label{eq:legmom2}
  \begin{aligned}
A_0 &= \frac{\sigma}{4\pi} = \frac{(\hbar c)^2}{32\pi s}
\frac{\omega(s,\mb_{a'}^2,\mm_{b'}^2)}{\omega(s,\mb_{a}^2,\mm_{b}^2)}
a_0~,\\
a_0\frac{A_1}{A_0} &= 2\Re\left(f_{0}(2f_{1+}^*+f_{1-}^*)\right) +
\frac{4}{5} \Re\left((2f_{1+}+f_{1-})(3f_{2+}^*+2f_{2-}^*)\right) \\
&+ \frac{12}{5} \Re\left((f_{1+}-f_{1-})(f_{2+}^*-f_{2-}^*)\right),\\
a_0\frac{A_2}{A_0} &= 2 \Re\left(f_{0}(3f_{2+}^*+2f_{2-}^*)\right) +
\frac{2}{3} \left|2f_{1+}+f_{1-}\right|^2 + \frac{2}{7}
\left|3f_{2+}+2f_{2-}\right|^2 - \frac{2}{3}
\left|f_{1+}-f_{1-}\right|^2 \\
&+ \frac{6}{7}
\left|f_{2+}-f_{2-}\right|^2~, \\
a_0\frac{B_1}{A_0} &= 2 \Im\left((f_{1+}^*-f_{1-}^*)f_{0}\right) -
\frac{2}{5} \Im\left((f_{1+}^*-f_{1-}^*)(3f_{2+}+2f_{2-})\right) \\
&+ \frac{6}{5} \Im\left((f_{2+}^*-f_{2-}^*)(2f_{1+}+f_{1-})\right)~,\\
a_0\frac{B_2}{A_0} &= 2 \Im\left(f_{1+}^*f_{1-}\right) + 2
\Im\left((f_{2+}^*-f_{2-}^*)f_{0}\right) + \frac{10}{7} \Im\left(f_{2+}^*f_{2-}\right)~.
  \end{aligned}
\end{equation}

The threshold branching fractions are defined as \cite{tov71,nov78},
\begin{equation}
  \label{eq:fracts}
  \gamma=\left.\frac{\sigma_{pK^-\rightarrow
      \Sigma^-\pi^+}}{\sigma_{pK^-\rightarrow \Sigma^+\pi^-}}\right|_\mathrm{thr} ~,
\quad
   R_c = \left.\frac{\sigma_{pK^-\rightarrow
       \mathrm{charged}}}{\sigma_{pK^-\rightarrow \mathrm{all}}}\right|_\mathrm{thr}~,
\quad
   R_n = \left.\frac{\sigma_{pK^-\rightarrow
       \Lambda\pi^0}}{\sigma_{pK^-\rightarrow \mathrm{all~neutral}}}\right|_\mathrm{thr}~.
\end{equation}
Finally, with our normalization the scattering length is expressed as,
\begin{equation}
\label{eq:scalen}
  a_{pK^-} = 
  \frac{\hbar c}{4\pi}\frac{1}{2m_p}\frac{1}{1+m_{K^-}/m_p}
  \left(\pw{0}{45}{45}\right)_\mathrm{thr}~,  
\end{equation}
with $(\pw{0}{45}{45})_\mathrm{thr}$ the $S$ wave for elastic $pK^-$
scattering evaluated at threshold.  Thus,
$(\sigma)_\mathrm{thr} = 4\pi |a_{pK^-}|^2$.

\newpage

\begin{figure}[h]
  \centering
\includegraphics{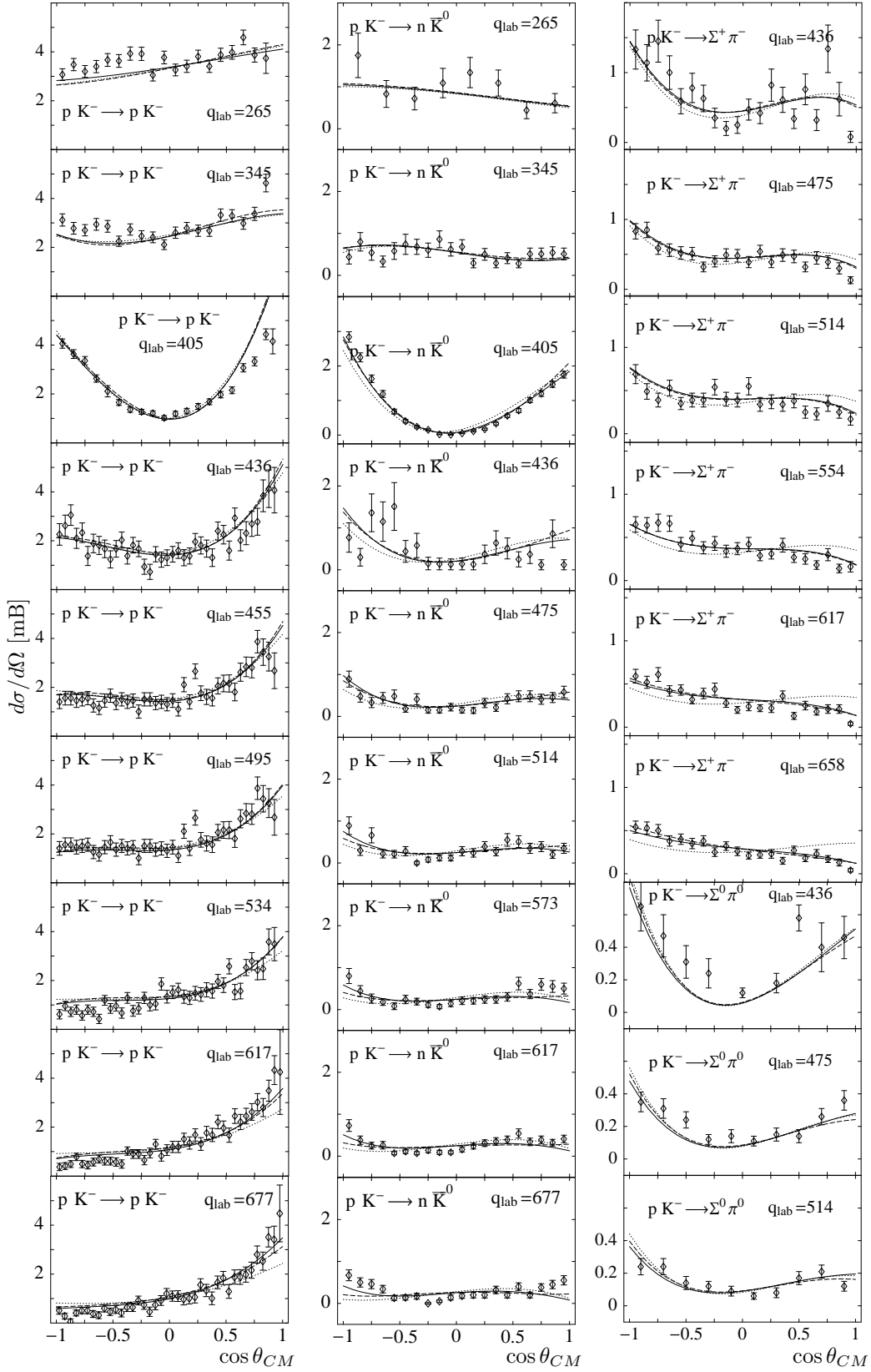}    
  \caption{Differential cross sections for $pK^-$ scattering. Solid,
    dashed and dotted lines as in fig.\ \ref{fig:1}. Only a
    representative sample of data from \cite{arm70,mas76} is shown.} 
  \label{fig:2}
\end{figure}

\addtocounter{figure}{-1}
\begin{figure}[h]
  \centering
\includegraphics{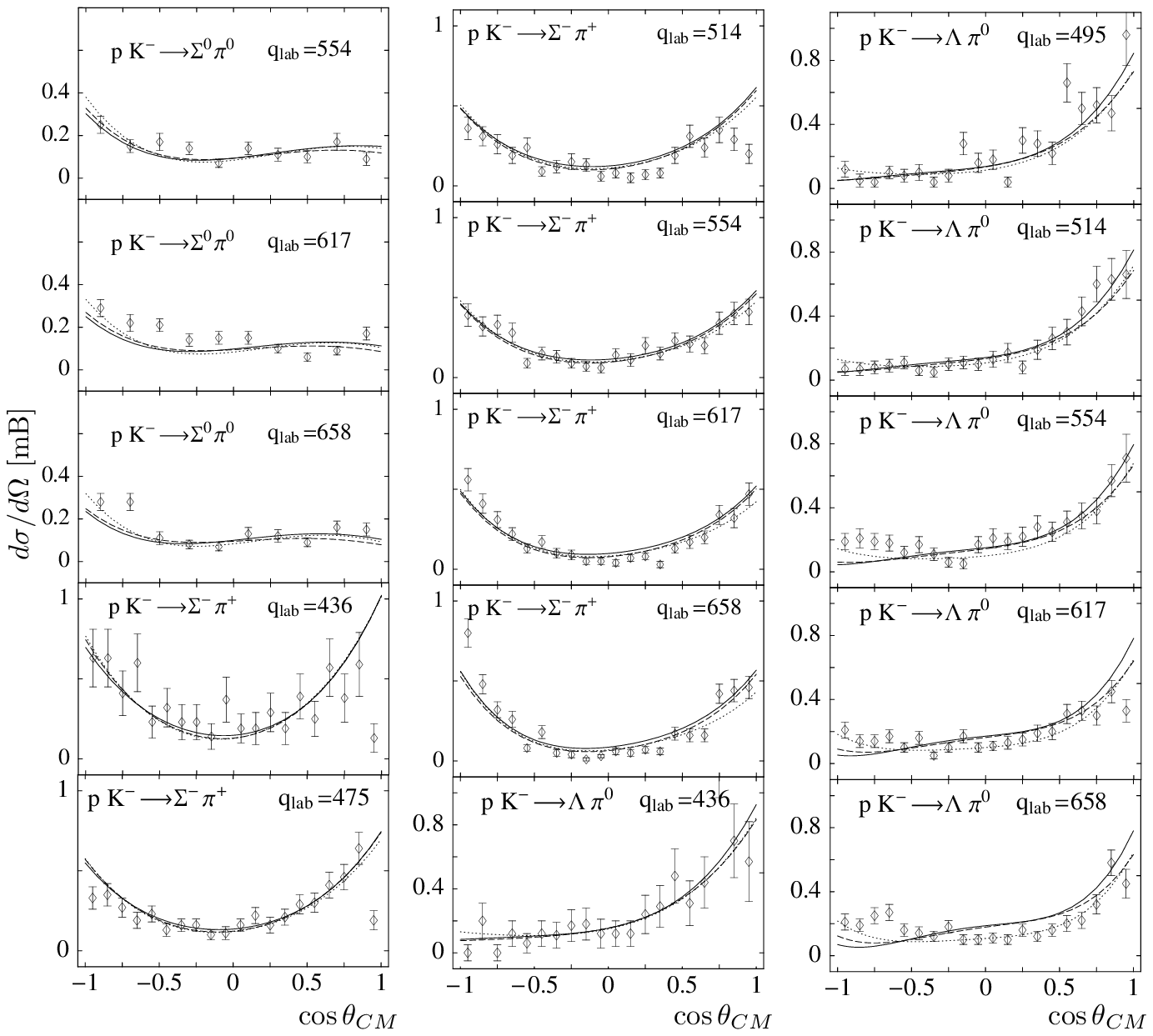}    
\caption{Continued.}
  \label{fig:2a}
\end{figure}

\begin{figure}[h]
  \centering
\includegraphics{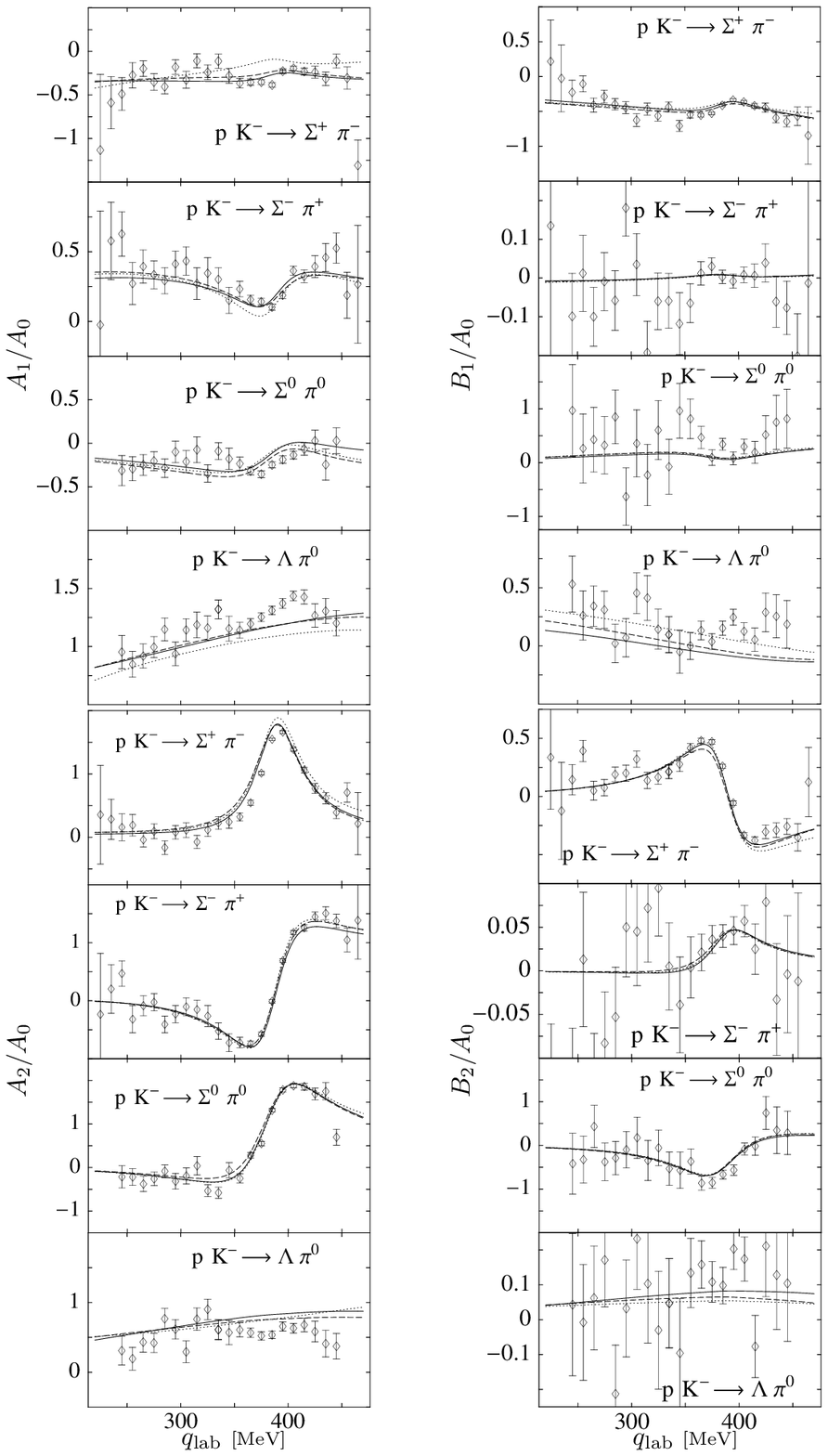}    
  \caption{Legendre moments of differential cross sections and spin
  asymmetries for $pK^-$ scattering. Solid, dashed and dotted lines as
  in fig.\ \ref{fig:1}. Data from \cite{mas75,ban81,cib82}.} 
  \label{fig:3}
\end{figure}

\begin{figure}[h]
  \centering
\includegraphics{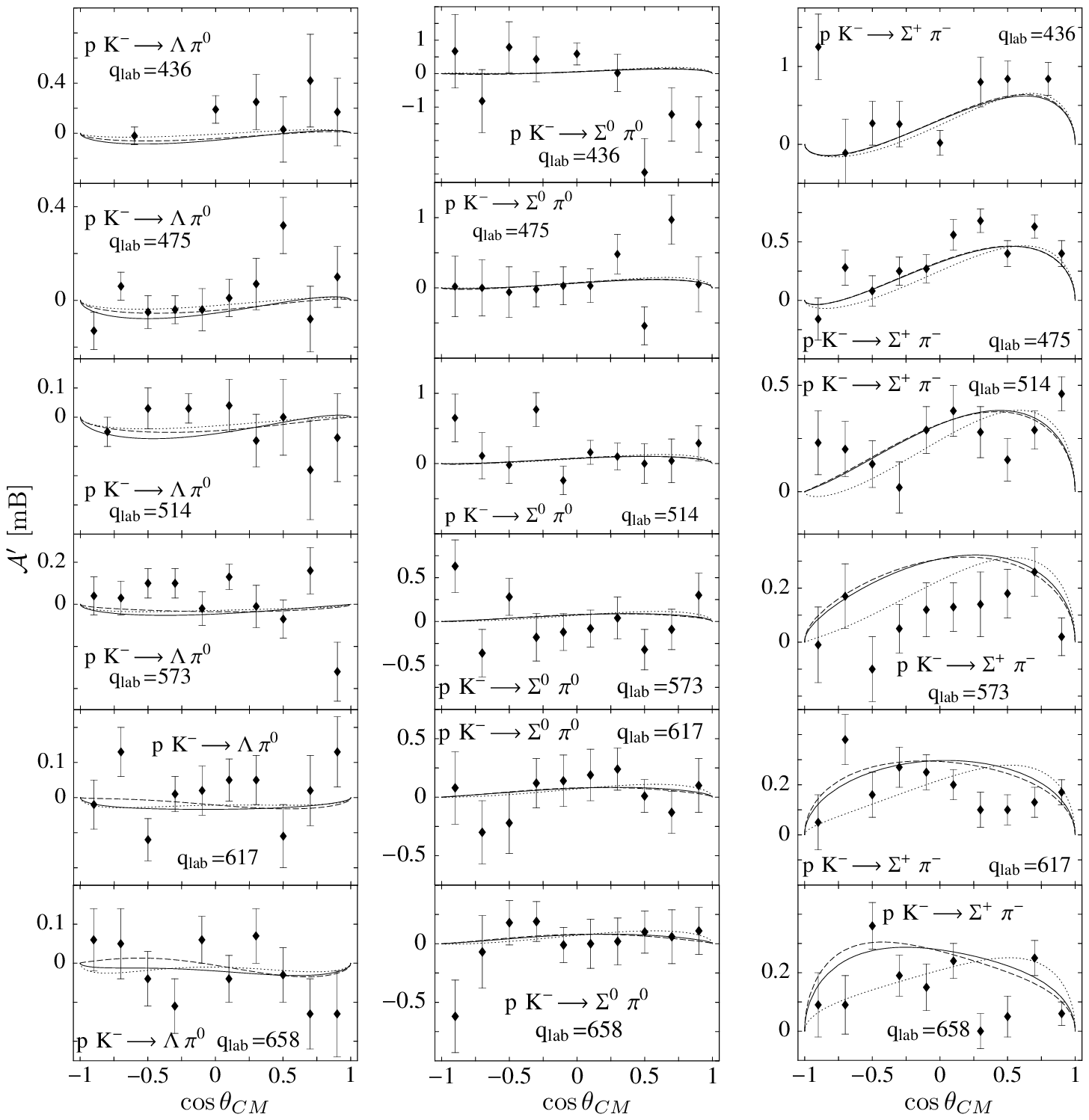}    
  \caption{Center-of-mass frame spin asymmetry for $pK^-$
    scattering. Solid, dashed and dotted lines as in fig.\
    \ref{fig:1}. Data from \cite{arm70}.} 
  \label{fig:4}
\end{figure}

\begin{figure}[h]
  \centering
\includegraphics{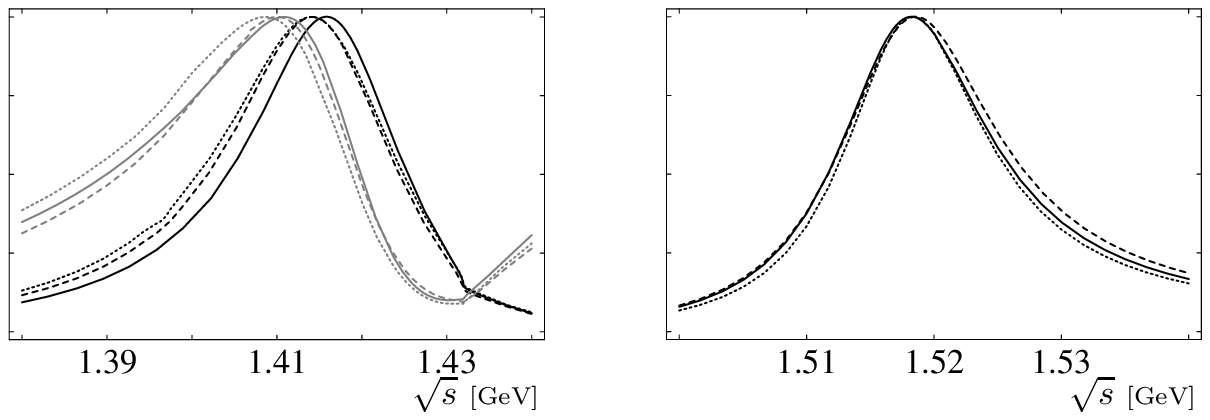}    
\caption{$\Sigma\pi$ mass spectrum for isosinglet partial waves, in
  arbitrary units.  Left panel: $S_{1/2}$ wave,
  $\Sigma\pi\rightarrow\Sigma\pi$ (gray lines) and
  $NK\rightarrow\Sigma\pi$ (black lines).  Right panel: $D_{3/2}$ wave,
  $NK\rightarrow\Sigma\pi$. Solid, dashed and dotted lines as in fig.\
  \ref{fig:1}.}
  \label{fig:5}
\end{figure}

\end{document}